\documentclass[preprint, preprintnumbers,showpacs, showkeys, aps, prd, amsmath, amsfonts, amssymb,
   eqsecnum, nofootinbib, floatfix, secnumarabic]{revtex4-1}

\usepackage{epsf}
\usepackage{epsfig}
\usepackage{graphics}
\usepackage{graphicx}
\usepackage{amssymb}

\def\gsim{\lower0.5ex\hbox{$\:\buildrel >\over\sim\:$}}
\def\lsim{\lower0.5ex\hbox{$\:\buildrel <\over\sim\:$}}

\begin{document}
\preprint{CUMQ/HEP 183}

\title{ \Large  Vector Quarks in the Higgs Triplet Model }

\author{Sahar Bahrami$^1$\footnote{Email: sahar.bahrami@concordia.ca}}
\author{Mariana Frank$^1$\footnote{Email: mariana.frank@concordia.ca}}

\affiliation{ $^1 $Department of Physics,  
Concordia University, 7141 Sherbrooke St. West ,
Montreal, Quebec, Canada H4B 1R6.}

\date{\today}

\begin{abstract}

We analyze the effects of introducing vector fermions in the Higgs Triplet Model. In this scenario, the model contains, in addition to the Standard Model particle content, one triplet Higgs representation, and a variety of vector-like fermion states, including singlet, doublet, and triplet states. We investigate the electroweak precision variables and impose restrictions on model parameters. We show that, for some representations, introducing vector quarks significantly alters the constraints on the mass of the doubly charged Higgs boson, bringing it in closer agreement with experimental constraints. We also study the effects of introducing the vector-like  fermions on neutral Higgs phenomenology, in particular on the loop-dominated decays $H \to \gamma \gamma$ and $H \to Z \gamma$, and the restrictions they impose on the parameter space.
\end{abstract}
\maketitle



 \section{Introduction}
  \label{sec:intro}
  
 The Standard Model (SM) of particle physics has received a big boost of confidence from the LHC Higgs data \cite{:2012gk}, as the discovery of the Higgs boson completes the model, and as the model appears so far to satisfy most, if not all, experimental constraints. Yet the SM fails to answer some fundamental questions, from both the theoretical and the experimental sides. Extensions of the SM resolve some of these questions, and while their predictions can overlap with the SM for phenomena where SM fits the experimental data, they can also resolve some conflicts of the SM with the data where such discrepancies exist. For instance, supplementing the SM by an additional complex Higgs triplet representation resolves naturally the origin of the neutrino mass \cite{Mohapatra:1980yp}, the existence of dark matter \cite{FileviezPerez:2008bj}, and provides an explanation for the excess in the Higgs decay into two photons \cite{Arbabifar:2012bd}.
 
 In addition to scalar fields, the SM can be extended by additional fermionic particles. 
 Some of the simplest extensions would include an additional  pair of chiral fermions, mimicking the already-existing fermion representations. However such models are all but ruled out by the Higgs data. An exception to this may be provided by including such additional representations in the Higgs Triplet Model (HTM) with non-trivial mixing between  the neutral CP-even Higgs states \cite{Banerjee:2013hxa}, but even there the parameter space is under significant pressure, and may be ruled out by data from the LHC operating at 13 TeV. The addition of non-chiral fermionic representations, such as vector quarks and/or vector leptons, is much less constrained. Vector-like fermions, which decay into SM fermions and a gauge boson or a Higgs particle, are predicted by extra-dimensional models \cite{Gopalakrishna:2013hua}, little Higgs models \cite{Han:2003wu}, heterotic string and string D-brane theories \cite{Dijkstra:2004cc} and by some composite Higgs models \cite{Contino:2006qr}. Vector-like fermions do not acquire mass through Yukawa couplings, they only affect the loop-dominated Higgs decay, and they may provide a better fit to the LHC Higgs data \cite{Bonne:2012im}. A great deal of literature is dedicated to analyses of vector fermions in the SM \cite{Carmi:2012yp,Ellis:2014dza,Aguilar-Saavedra:2013qpa}, as well as in model-independent scenarios \cite{Moreau:2012da}.
 
 In general, fewer studies involve introducing vector fermions into specific non-SM models.  Supplementing these models by additional vector fermion states can alleviate some of the  restrictions on the parameters in these scenarios. For instance, adding vector leptons in the two Higgs doublet model \cite{Garg:2013rba}  alleviates electroweak precision constraints. In supersymmetry,  vector leptons can improve vacuum stability and enhance the di-photon rate by as much as 50\% \cite{Joglekar:2013zya}.
  
In a previous work \cite{Bahrami:2013bsa} we showed that, if light enough, vector leptons introduced into the Higgs Triplet Model modify {\it both} the decay rates of the neutral Higgs boson into two photons, {\it and} the decay patterns and branching ratios of the doubly charged Higgs bosons.  In this work, we extend our study to a carefully general consideration of the theoretical and phenomenological implications of additional vector-like quarks  states in the HTM. 
The effects of the vector quarks in the Higgs Triplet Model  on the Higgs decays has been investigated before in \cite{Wang:2012gm}, where the authors showed that for some values of the couplings between the Higgs boson and the vector-like quarks,  the decay $H \to \gamma \gamma$ can be enhanced. Our approach here is very different than theirs. We specify the possible hypercharge assignments for the new quarks, and then allow their masses and couplings to be free parameters. We study cases in which  vector-like states couple to the gauge fields and mix weakly with SM quarks of the third generation only, to avoid flavor violation problems.  We investigate the precision electroweak constraints due to their presence in the HTM and the impact of vector-like states on the Higgs branching fractions, particularly into two photons and into Z$\gamma$. Unlike vector leptons, vector-like quarks affect {\it both} the production cross section and the decay rates of the Higgs bosons. We present numerical results which restrict the masses and mixings of the new vector-like quarks, and which  have implications for future vector fermion searches. We also revisit the implications of their inclusion for doubly charged Higgs states. 

Our work is organized as follows. In the next section Sec. \ref{sec:model} we summarize the basics features of the Higgs Triplet Model without (in \ref{subsec:HTM})  and with (in \ref{subsec:VQ} )vector-like quarks. We define the representations, as well as masses and mixing parameters. We proceed by examining the electroweak precision constraints in Sec. \ref{sec:STU} in the HTM, again without (\ref{subsec:THTM}) and with (\ref{subsec:TSVQ}) vector-like quarks. In the same section, we present a numerical analysis on the restrictions coming from the oblique parameters on the masses of the doubly charged Higgs bosons, and on the masses and mixing parameters with third generation quarks for the vector-like quarks, in \ref{subsec:mHpp}. These restrictions are then applied to evaluation of the relative (with respect to the SM) branching decay rates for $H\to \gamma \gamma$ and $H \to Z \gamma$ in Sec. \ref{sec:hgg}. We summarize our findings and conclude in Sec. \ref{sec:conclusion}. Some definition of our parameters are included in the Appendix \ref{sec:appendix}.
\section{The Model}
\label{sec:model}
\subsection{Higgs Triplet Model}
\label{subsec:HTM}
The Higgs Triplet Model (HTM) has been studied extensively in \cite{Arhrib:2011uy,Kanemura:2012rs}. The symmetry group is the same as that in the SM, $SU(2)_L \times U(1)_Y$, but  one triplet field $\Delta$ with hypercharge $Y=1$ is added to the SM Higgs sector, which already contains one isospin doublet field $\Phi$ with 
hypercharge $Y=1/2$. The Higgs fields are given by:
\begin{eqnarray}
\Phi=\left[
\begin{array}{c}
\varphi^+\\
\frac{1}{\sqrt{2}}(\varphi+v_\Phi+i\chi)
\end{array}\right],\quad \Delta =
\left[
\begin{array}{cc}
\frac{\Delta^+}{\sqrt{2}} & \Delta^{++}\\
\frac{1}{\sqrt{2}}(\delta+v_\Delta+i\eta) & -\frac{\Delta^+}{\sqrt{2}} 
\end{array}\right],
\end{eqnarray}
where $v_\Phi$ and $v_\Delta$ 
are the VEVs of the doublet Higgs field and the triplet Higgs field, with 
$v^2\equiv v_\Phi^2+2v_\Delta^2\simeq$ (246 GeV)$^2$.  
The Higgs potential involving the doublet $\Phi$ and triplet $\Delta$  is 
\begin{eqnarray}
V(\Phi,\Delta)&=&m^2\Phi^\dagger\Phi+M_t^2\rm{Tr}(\Delta^\dagger\Delta)+\left[\mu \Phi^Ti\tau_2\Delta^\dagger \Phi+\rm{h.c.}\right]+\lambda_1(\Phi^\dagger\Phi)^2 \nonumber\\
&+&\lambda_2\left[\rm{Tr}(\Delta^\dagger\Delta)\right]^2 +\lambda_3\rm{Tr}[(\Delta^\dagger\Delta)^2]
+\lambda_4(\Phi^\dagger\Phi)\rm{Tr}(\Delta^\dagger\Delta)+\lambda_5\Phi^\dagger\Delta\Delta^\dagger\Phi,~~~~ 
\label{eq:pot_htm}
\end{eqnarray}
with parameters (all assumed real), $m$ and $M_t$ the Higgs bare masses, $\mu$ the lepton-number violating  parameter,    and 
$\lambda_1$-$\lambda_5$, the Higgs coupling constants. 
The scalar potential in Eq. (\ref{eq:pot_htm}) induces mixing among the physical states for the singly charged, the CP-odd, and the CP-even neutral scalar sectors, respectively:
\begin{eqnarray}
\left(
\begin{array}{c}
\varphi^\pm\\
\Delta^\pm
\end{array}\right)&=&
\left(
\begin{array}{cc}
\cos \beta_\pm & -\sin\beta_\pm \\
\sin\beta_\pm   & \cos\beta_\pm
\end{array}
\right)
\left(
\begin{array}{c}
w^\pm\\
H^\pm
\end{array}\right),\quad 
\left(
\begin{array}{c}
\chi\\
\eta
\end{array}\right)=
\left(
\begin{array}{cc}
\cos \beta_0 & -\sin\beta_0 \\
\sin\beta_0   & \cos\beta_0
\end{array}
\right)
\left(
\begin{array}{c}
z\\
A
\end{array}\right),\nonumber\\
\left(
\begin{array}{c}
\varphi\\
\delta
\end{array}\right)&=&
\left(
\begin{array}{cc}
\cos \alpha & -\sin\alpha \\
\sin\alpha   & \cos\alpha
\end{array}
\right)
\left(
\begin{array}{c}
h\\
H
\end{array}\right),
\end{eqnarray}
with mixing angles  given by  
\begin{eqnarray}
\tan\beta_\pm&=&\frac{\sqrt{2}v_\Delta}{v_\Phi},\quad \tan\beta_0 = \frac{2v_\Delta}{v_\Phi}, \nonumber\\
\tan2\alpha &=&\frac{v_\Delta}{v_\Phi}\frac{2v_\Phi^2(\lambda_4+\lambda_5)-4M_\Delta^2}{2v_\Phi^2\lambda_1-M_\Delta^2-2 v_\Delta^2(\lambda_2+\lambda_3)}.~~~~~~~~ \label{tan2a}
\end{eqnarray}
The CP-even Higgs states which mix with the angle $\alpha$ are given, in terms of the couplings in the scalar potential, by
\begin{eqnarray}
&&m_h^2= 2v_\Phi^2\lambda_1\cos^2\alpha+\left [M_\Delta^2+2v_\Delta^2(\lambda_2+\lambda_3)\right ] \sin^2\alpha+\left [\frac{2v_\Delta}{v_\Phi}M_\Delta^2-v_\Phi v_\Delta(\lambda_4+\lambda_5)\right] \sin2\alpha,\nonumber\\
&&\\
&&m_H^2=2v_\Phi^2\lambda_1\sin^2\alpha+ \left [M_\Delta^2+2v_\Delta^2(\lambda_2+\lambda_3)\right] \cos^2\alpha- \left [\frac{2v_\Delta}{v_\Phi}M_\Delta^2-v_\Phi v_\Delta(\lambda_4+\lambda_5)\right ] \sin2\alpha,\nonumber\\
\end{eqnarray}
where we defined  $\displaystyle M_\Delta^2\equiv \frac{v_\Phi^2\mu}{\sqrt{2}v_\Delta^2}$. 
Note that, while the mixing angles in the charged and CP-odd sectors are constrained to be small by the hierarchy of the VEVs, the same is not necessarily  the case for $\alpha$. In fact, as we have previously shown, if and only if $\alpha$ is allowed to be non-zero, yielding significant mixing in the CP-even neutral sector, the decay of one of the neutral Higgs bosons into two photons can be enhanced \cite{Arbabifar:2012bd}. 
The parameters of the model are restricted by the values of the  $W$ and  $Z$  masses and the electroweak $\rho$ parameter, defined at tree level  
\begin{eqnarray}
m_W^2& =& \frac{g^2}{4}(v_\Phi^2+2v_\Delta^2),\quad m_Z^2 =\frac{g^2}{4\cos^2\theta_W}(v_\Phi^2+4v_\Delta^2), \nonumber \\
\rho &\equiv & \frac{m_W^2}{m_Z^2\cos^2\theta_W}=\frac{1+\frac{2v_\Delta^2}{v_\Phi^2}}{1+\frac{4v_\Delta^2}{v_\Phi^2}},  \label{rho_triplet}
\end{eqnarray}
insuring the smallness of $v_\Delta/v_\Phi$. The parameters of the model are further restricted by the  smallness of the  Majorana neutrino masses, proportional to the lepton number violating coupling constant $\mu$
\begin{equation}
(m_\nu)_{ij}=\sqrt{2}h_{ij} v_\Delta=h_{ij}\frac{\mu v_\Phi^2}{M_\Delta^2}, \label{mn}
\end{equation}
requiring $\mu\ll M_\Delta$ for the smallness of the neutrino masses to be explained by the type II seesaw mechanism. 

In a previous work \cite{Bahrami:2013bsa} we have shown that introducing vector-like leptons in the model can significantly alter the decay patterns of the doubly charged Higgs bosons and thus modify the experimental bounds on their masses. We adopt here the same model parameters, and allow $\sin \alpha$ to vary, set $m_h=125$ GeV and $m_H=98$ GeV.  We proceed by introducing vector-like quarks and study their effects in the HTM.
\subsection{Higgs Triplet Model with Vector-Like Quarks}
\label{subsec:VQ}
In considering addition of vector leptons to the Higgs Triplet Model, the representations
considered included $SU(2)_L$ lepton doublets, right-handed charged and neutral vector singlets and their mirror images. Our assumption was that the vector-like leptons can be light, and then introduced a parity symmetry which forbade mixing between the new vector-like fields (odd under this symmetry) and  the ordinary leptons (even under the same symmetry). This insured that flavor, stringently constrained in ordinary lepton decays, was not violated. 

Introduction of vector-like quarks imposes different constraints on the HTM, and thus the scenarios presented here would be qualitatively different from those introduced in  \cite{Bahrami:2013bsa}. 
First, vector-like quarks affect both the production and decay of the Higgs bosons at the LHC.  Second, flavor violation is less constrained in the quark sector, allowing the new vector-like states to mix weakly with the third family of ordinary quarks.  In this subsection we introduce vector-like quarks into the model, and in the next section we study their effects. 
We first classify the vector-like quarks in terms of multiplets of $SU(2)_L \times U(1)_Y$, then proceed by writing gauge invariant interactions for each. The new states interact with the Higgs states through Yukawa interactions. The allowed multiplet states for the  vector-like quarks, together with their nomenclature, are listed in Table \ref{tab:VQrepresentations} \cite{Carmi:2012yp,Ellis:2014dza,Aguilar-Saavedra:2013qpa}. The first two representations are $U$-like and $D$-like singlets, the next three are doublets (one SM-like, two non-SM like), and the last two are triplets. The various representations are distinguished by their $SU(2)_L$ and hypercharge numbers.
\begin{table}[htbp]
\caption{\label{tab:VQrepresentations}\sl\small Representations of Vector-Like Quarks, with quantum numbers under $SU(2)_L \times U(1)_Y$.}
  \begin{center}
 \small
 \begin{tabular*}{0.99\textwidth}{@{\extracolsep{\fill}} c| ccccccc}
 \hline\hline
	Name &${\cal U}_1$ &${\cal D}_1$ &${\cal D}_2$ &${\cal D}_X$ &${\cal D}_Y$ &${\cal T}_X$ 
	&${\cal T}_Y$\\
  Type&Singlet &Singlet &Doublet&Doublet &Doublet &Triplet 
	&Triplet\\
	 \hline
	   &$T$ &$B$ &$\left ( \begin{array}{c} T \\ B \end{array} \right ) $ &$ \left (\begin{array}{c}  X \\ T  \end{array}\right)$ &$\left ( \begin{array}{c} B \\ Y\end{array} \right ) $ &$ \left (\begin{array}{c} X\\T \\ B \end{array} \right ) $
	  & $\left ( \begin{array}{c} T \\ B\\Y \end{array} \right )$\\
  \hline
  $SU(2)_L$ &1 &1 &2 &2 & 2 &3 &3 \\
  \hline
  $Y$ &$ 2/3$ &$ -1/3$ &$1/6 $ &$ 7/6$ &$-5/6$ &$2/3$ &$-1/3$ \\
      \hline
    \hline
   \end{tabular*}
\end{center}
 \end{table}

In these representations, Yukawa and the relevant interaction terms  between the vector-like quarks and SM quarks are \cite{Cai:2012ji}
\begin{eqnarray}
{\cal L}_{SM}&=& -y_u {\bar q}_LH^c u_R -y_d{\bar q}_L H d_R \nonumber \\
{\cal L}_{{\cal U}_1, {\cal D}_1}&=& -\lambda_u {\bar q}_LH^c U_{1_R} -\lambda_d{\bar q}_L H D_{1_R}-M {\bar U}_L U_R-M {\bar D}_L D_R, \nonumber \\
{\cal L}_{{\cal D}_2}~~&=& -\lambda_u {\bar D}_{2_L} H^c u_{R} -\lambda_d{\bar D}_{2_L} H d_{R}-M {\bar D}_{2_L}  D_{2_R}, \nonumber \\
{\cal L}_{{\cal D}_X, {\cal D}_Y}&=& -\lambda_u {\bar D}_{X_L}H u_{R} -\lambda_d{\bar D}_{Y_L} H^c d_{R}-M {\bar D}_{X_L} D_{X_R}-M {\bar D}_{Y_L} D_{Y_R}, \nonumber \\
{\cal L}_{{\cal T}_X, {\cal T}_Y}&=& -\lambda_u {\bar q}_{L}\tau^a H^c  {\cal T}^a_{X_R} -\lambda_d{\bar q}_{L} \tau^a H {\cal T}^a_{Y_R}-M {\bar  {\cal T}}_{X_L}  {\cal T}_{X_R}-M {\bar  {\cal T}}_{Y_L}  {\cal T}_{Y_R}.
\end{eqnarray}
After the spontaneous symmetry breaking, the Yukawa interactions generate mixing between the SM quarks and the vector quarks at tree level. The singlet vector-like quark and the triplet vector-like quark exhibit  similar mixing patterns, while the doublet vector-like quark has a different mixing pattern \cite{Cai:2012ji}. To avoid conflicts with low energy  experimental data, we consider that  the vector-like quarks mix with the third generation of SM quarks only.

The mass matrix for the mixing between $m_t$ and $m_{T}$ can be diagonalized by two mixing matrices:
\begin{eqnarray}
V_L^u= \left (\begin{array}{cc} \cos \theta_L^u & \sin \theta_L^u\\
-\sin \theta_L^u& \cos \theta_L^u \end{array} \right ), \qquad  V_R^u= \left (\begin{array}{cc} \cos \theta_R^u & \sin \theta_R^u\\
-\sin \theta_R^u& \cos \theta_R^u \end{array} \right ), 
\end{eqnarray}
for the singlet/triplet vector quark, such that
\begin{equation}
\left (\begin{array}{cc} \cos \theta_L^u & -\sin \theta_L^u\\
\sin \theta_L^u& \cos \theta_L^u \end{array} \right ) \left (\begin{array}{cc} \frac{y_u v}{\sqrt{2}} & x_t\\
0& M \end{array} \right ) \left (\begin{array}{cc} \cos \theta_R^u & \sin \theta_R^u\\
-\sin \theta_R^u& \cos \theta_R^u \end{array} \right ) =\left (\begin{array}{cc} m_t & 0\\
0& m_{T} \end{array} \right ),
\label{eq:mixing_s,t}
\end{equation}
where $m_{T} \ge M \ge m_t$. Similar relations hold for $m_b$ and $m_{B}$. The relations between the tree-level  input parameters and the mixing angles and masses are given by \cite{Cacciapaglia:2010vn}:
\begin{eqnarray}
\frac{y^2_u v^2}{2}&=&m_t^2\left ( 1+\frac{x_t^2}{M^2-m_t^2} \right ) \nonumber\\
m_{T}^2&=&M^2\left ( 1+\frac{x_t^2}{M^2-m_t^2} \right ), \nonumber\\
\sin \theta^{u,d}_L&=&\frac{Mx_{t(b)}}{\sqrt{(M^2-m_{t(b)}^2)^2+M^2x_{t(b)}^2}}\nonumber\\
\sin \theta^{u,d}_R&=&\frac{m_{t(b)} x_{t(b)}}{\sqrt{(M^2-m_{t(b)}^2)^2+M^2 x_{t(b)}^2}},
\label{eq:eigenvalues}
\end{eqnarray}
where $\displaystyle x_t=\frac{\lambda_u v}{\sqrt{2}}$ and $\displaystyle x_b=\frac{\lambda_d v}{\sqrt{2}}$.
For the case of doublets, the diagonalization can be carried out in a similar way:
\begin{equation}
\left (\begin{array}{cc} \cos \theta_L^u & -\sin \theta_L^u\\
\sin \theta_L^u& \cos \theta_L^u \end{array} \right ) \left (\begin{array}{cc} \frac{y_u v}{\sqrt{2}} & 0\\
x& M \end{array} \right ) \left (\begin{array}{cc} \cos \theta_R^u & \sin \theta_R^u\\
-\sin \theta_R^u& \cos \theta_R^u \end{array} \right ) =\left (\begin{array}{cc} m_t & 0\\
0& m_{T} \end{array} \right ).
\label{eq:mixing_d}
\end{equation}
The relations between the parameters are the same, except that the formulas for the left- and right-handed mixing angles are interchanged:
\begin{eqnarray}
\sin \theta^{u,d}_L&=&\frac{m_{t(b)} x_{t(b)}}{\sqrt{(M^2-m_{t(b)}^2)^2+M^2 x_{t(b)}^2}},\nonumber \\
\sin \theta^{u,d}_R&=&\frac{Mx_{t(b)}}{\sqrt{(M^2-m_{t(b)}^2)^2+M^2x_{t(b)}^2}} .
\end{eqnarray}
We use the shorthand notations $s^{u,d}_L \equiv\sin \theta^{u,d}_L$ and $c^{u,d}_L \equiv\cos \theta^{u,d}_L$. Note that in the ${\cal T}_X$ triplet model, the two mixing angles are related to each other by $x_b= \sqrt{2}  x_t$. In the ${\cal T}_Y$ model, for bottom sector $x_b=-x_t$ and for the the top the same formulas as in other case apply, with $x_t\rightarrow \sqrt{2} x_t$ \cite{Cai:2012ji,Cacciapaglia:2010vn}. All multiplets thus involve at least one mixing angle. These mixed states will be used to express interactions with the Higgs and gauge bosons, and constrain those interactions. 
The mixing of a $b$ quark with a heavy vector-like $B$ quark modifies the $Z b{\bar b}$ coupling at the tree level, while the mixing between a $t$ quark with a heavy vector-like $T$ modifies the $W b{\bar t}$ vertex.  We compute both of these, using analytical expressions.
In ${\cal D}_1$ model\footnote{These corrections are scenario-dependent. More general formulas have appeared elsewhere \cite{Cacciapaglia:2010vn}.}, the strongest tree level bound comes from correction to $Z b_l{\bar b_l}$ coupling:
\begin{eqnarray}
\delta R_b=2R_b(1-R_b)\frac{\delta g_{ZbL}}{\delta g^{SM}_{ZbL}}, 
\end{eqnarray}
where
\begin{eqnarray}
\delta g^{SM}_{ZbL} = 1-\frac{2}{3}\sin^2 \theta_W, \qquad
\delta g_{ZbL} = s_L^{d\,2}.
\end{eqnarray}
Here $g^{SM}_{ZbL}$ is $Z$-boson coupling to the left-handed $b$ quark in the SM,  $R_b$ is defined as $\displaystyle \frac {\Gamma ( Z\rightarrow b {\bar b})} {\Gamma ( Z\rightarrow \rm hadrons)}$,  with its SM value $R_b= 0.21578_{-0.0008}^{+0.0005}$ \cite{Cai:2012ji}. Electroweak measurements constraints for the deviation $\delta R_b$ due to the new physics effects are  $\delta R_b= 0.00051\pm 0.00066$ \cite{Beringer:1900zz}, and experimental restrictions \cite{ALEPH:2005ab} are  $[Zb {\bar b}]_{\rm exp}=0.21629 \pm 0.00066$. The relevant couplings for the models analyzed are included in the Appendix.

In ${\cal D}_X$ model, the tree level bound comes from the left-handed $W b{\bar t}$ coupling: 
\begin{eqnarray}
\frac{\delta g_{W}}{\delta g^{SM}_{W}}= c_L^{u}-1.
\end{eqnarray} 
Experimental searches for vector-like quarks have set mass limits on some of the representations.  Current experimental bounds depend critically on the details of the models and assumptions about branching ratios. 
A lot of searches at  ATLAS and CMS  focused on a top-like quark with charge $+2/3$.  The mass limits obtained are $m_T>640$ GeV, for ${\cal U}_1$, ${\cal D}_2$ and  ${\cal T}_X$ models \cite{ATLAS:2013t,ATLAS:2012qe}, and $m_T>790$ GeV, for ${\cal D}_X$ and ${\cal T}_Y$ models \cite{ATLAS:2013t}. Some bounds exist for bottom-like vector quarks, and the bounds obtained are $m_B>590$ GeV, for ${\cal D}_1$ and ${\cal T}_Y$ models \cite{ATLAS:2013b}, and $m_B>358$ GeV, for ${\cal D}_2$ and ${\cal D}_Y$ models \cite{Aad:2012pga}. Mass limits also exist for the exotic X quarks: $m_X>770$ GeV, in ${\cal D}_X$ and ${\cal T}_X$ models \cite{CMS:2013x}, and for the Y quark $m_Y>656$ GeV, in ${\cal D}_Y$ and ${\cal T}_Y$ models \cite{ATLAS:2012qe}. 
However, the vector-like quarks could have escaped detection so far by prompt decays, and even relaxed limits on the mixing between top and vector-like top quarks can avoid the present experimental bounds  \cite{Ellis:2014dza}.


\section{Electroweak Constraints}
\label{sec:STU}
The Peskin-Takeuchi parameters $S, T$ and $U$ are commonly used to constrain and characterize new physics, as a means to comparing its predictions with the electroweak precision data. They can be calculated perturbatively in any model from the gauge boson propagator functions, and are defined as \cite{Hagiwara:1994pw}: 
\begin{eqnarray}
S &=& 16\pi {\rm Re} \left [ {\bar \Pi}^{3Q}_{T,\gamma}(m_Z^2)-{\bar \Pi}^{33}_{T,Z}(0) \right], \nonumber\\
T &=& \frac{4 \sqrt{2} G_F}{\alpha_e} {\rm Re} \left [ {\bar \Pi}_T^{33}(0)-{\bar \Pi}^{11}_{T}(0) \right], \nonumber \\
U &=& 16\pi {\rm Re} \left [ {\bar \Pi}^{33}_{T,Z}(0)-{\bar \Pi}^{11}_{T,W}(0) \right],  
\end{eqnarray}
where the gauge boson two-point functions are defined as 
$ \displaystyle
{\bar \Pi}^{AB}_{T,V}(p^2) = \frac{{\bar \Pi}^{AB}_{T}(p^2)-{\bar \Pi}^{AB}_{T}(m_V^2)}{p^2-m_V^2}$,  
 and $\alpha_e\equiv \alpha_e(m_Z^2)$. The current experimental bounds defining $\Delta T=T-T_{\rm SM}, \Delta S=S-S_{\rm SM}$ and considering $\Delta U=0$, are 
$\Delta S= 0.05 \pm 0.09,~
\Delta T= 0.08 \pm 0.07$ \cite{Kanemura:2012rs} .  In our considerations  we allow for a more conservative deviation for the $\Delta T$ parameter between $-0.2$ and $0.4$ \cite{Cacciapaglia:2010vn}.

\subsection{Contributions to the $S, T$ and $U$-parameters in the HTM}
\label{subsec:THTM}
The explicit expressions for the $S$, $T$ and $U$ parameters for the HTM, including the extra Higgs representation, but without the vector-like quarks, are
\begin{eqnarray}
S_{HTM} &=& 16\pi {\rm Re} \left [ {\Pi}^{3Q}_{HTM,\gamma}(m_Z^2)-{\Pi}^{33}_{HTM,Z}(0) \right], \nonumber\\
T_{HTM} &=& \frac{4 \sqrt{2} G_F}{\alpha_e} {\rm Re} \left [ {\Pi}_{HTM}^{33}(0)-{\Pi}^{11}_{HTM}(0) \right], \nonumber \\
U_{HTM}&=& 16\pi {\rm Re} \left [ {\Pi}^{33}_{HTM,Z}(0)-{\Pi}^{11}_{HTM,W}(0) \right].  
\end{eqnarray}
where $\Pi^{AB}_{\rm HTM}(p^2)$ are the gauge boson two-point functions in the Higgs Triplet Model.The coupling factors are
$\displaystyle {\hat g}_Z=\frac {{\hat g}}{\cos \theta_W}$, so 
\begin{eqnarray}
\Pi^{3Q}_{\rm HTM}(p^2)&=&\frac{\Pi^{Z\gamma}(p^2)}{\sin \theta_W \cos \theta_W{\hat g}_Z^2} + \frac{\Pi^{\gamma\gamma}(p^2)}{\cos^2 \theta_W {\hat g}_Z^2},  \nonumber \\
\Pi^{33}_{\rm HTM}(p^2)&=&\frac{\Pi^{ZZ}(p^2)}{{\hat g}_Z^2} + \frac{2 \sin \theta_W \Pi^{Z\gamma}(p^2)}{\cos \theta_W {\hat g}_Z^2} + \frac{\sin^2 \theta_W \Pi^{\gamma \gamma }(p^2)}{\cos^2 \theta_W {\hat g}_Z^2}, \nonumber \\
\Pi^{11}_{\rm HTM}(p^2)&=&\frac{1} {\cos^2 \theta_W {\hat g}_Z^2} \Pi_{\rm HTM}^{WW}, 
\end{eqnarray}
evaluated at physical momentum transfers scales $\displaystyle p^2=0, m_Z^2, m_W^2$.

 In the HTM with and without vector-like quarks, the $S$ parameter is far less restricted by the parameters of the model, and does not pose difficulties in any of the models listed in Table \ref{tab:VQrepresentations}. While we shall plot the dependence of both $S$ and $T$ parameters on the variables of the HTM, we  give the explicit results for the $T$ parameter only.

The $W$-boson two-point function in the HTM is  \cite{Kanemura:2012rs} : 
\begin{eqnarray}
\Pi_{\rm HTM}^{WW}(p^2)&=&\frac{g^2}{16 \pi^2}\bigg \{ g^2\Big [ \left (\frac{v_\phi}{2}c_\alpha+v_\Delta s_\alpha \right )^2 B_0(p^2,m_h,m_W)+
\left (-\frac{v_\phi}{2}s_\alpha+v_\Delta c_\alpha \right )^2 B_0(p^2,m_H,m_W)\nonumber \\
&+&2 v_\Delta^2 B_0(p^2, m_{H^{\pm \pm}}, m_W)+ \frac{c^2_{\beta_\pm}}{2 c_W^2} v_\Delta^2 B_0(p^2, m_{H^{\pm}}, m_W) \nonumber \\
&+&\frac{1}{c_W^2}\left [\frac{v_\phi}{2} s_W^2c_{\beta_\pm}+\frac{v_\Delta}{\sqrt{2}}(1+s_W^2) s_{\beta_\pm} \right]^2B_0(p^2,m_Z,m_W)
\Big ]+\frac{e^2}{4}(v_\phi^2+2v_\Delta^2)B_0(p^2,0,m_W) \nonumber \\
&+& \Big [ \frac14 \Big [( c_\alpha s_{\beta_\pm}-\sqrt{2} s_\alpha c_{\beta_\pm})^2 B_5(p^2,m_{H^{\pm}}, m_{h})
+ (c_\alpha c_{\beta_\pm}+\sqrt{2} s_\alpha s_{\beta_\pm})^2 B_5(p^2,m_{W}, m_{h}) \nonumber \\
&+&(s_\alpha s_{\beta_\pm} +\sqrt{2}c_\alpha c_{\beta_\pm})^2 B_5(p^2,m_{H^\pm}, m_H)
+ (s_\alpha c_{\beta_\pm} -\sqrt{2}c_\alpha s_{\beta_\pm})^2 B_5(p^2,m_{W}, m_H)\nonumber \\
&+&(s_{\beta_0} s_{\beta_\pm} +\sqrt{2}c_{\beta_0} c_{\beta_\pm})^2 B_5(p^2,m_{H^\pm}, m_A)
+ (s_{\beta_0} c_{\beta_\pm} -\sqrt{2}c_{\beta_0} s_{\beta_\pm})^2 B_5(p^2,m_{W}, m_A)\nonumber \\
&+&(-c_{\beta_0} s_{\beta_\pm} +\sqrt{2}s_{\beta_0} c_{\beta_\pm})^2 B_5(p^2,m_{H^\pm}, m_Z)
+(c_{\beta_0} c_{\beta_\pm} +\sqrt{2}s_{\beta_0} s_{\beta_\pm})^2 B_5(p^2,m_{W}, m_Z)\Big ]\nonumber \\
&+& c_{\beta_\pm}^2 B_5(p^2,m_{H^{\pm\pm}}, m_{H^\pm})+s_{\beta_\pm}^2 B_5(p^2,m_{H^{\pm\pm}}, m_{W})
\bigg \}.
\end{eqnarray}
The photon two-point function is calculated as:
\begin{eqnarray}
\Pi_{\rm HTM}^{\gamma \gamma}(p^2)&=&\frac{e^2}{16 \pi^2} \Big [\frac{g^2}{2}(v_\phi^2+2v_\Delta^2)B_0(p^2, m_W, m_W)
+4B_5(p^2, m_{H^{\pm\pm}}, m_{H^{\pm\pm}})\nonumber \\
&+&B_5(p^2, m_{H^{\pm}}, m_{H^{\pm}}) +B_5(p^2, m_{W}, m_{W}) \Big].
\end{eqnarray}
The $Z$-boson two-point function in the HTM is 
\begin{eqnarray}
\Pi_{\rm HTM}^{ZZ}(p^2)&=& \frac{g_Z^2}{16 \pi^2} \Big \{m_Z^2 \left [ (c_{\beta_0} c_\alpha +2 s_{\beta_0} s_\alpha )^2 B_0(p^2,m_h,m_Z)+ (c_{\beta_0} s_\alpha -2 s_{\beta_0} c_\alpha )^2 B_0(p^2,m_H , m_Z)\right ]\nonumber \\
&+& m_W^2 \left [ 2 c_{\beta_\pm}^2 s_{\beta_\pm}^2 B_0(p^2, m_{H^\pm}, m_W) +2(s_W^2 +s_{\beta_\pm}^2)^2B_0(p^2, m_{G^\pm}, m_W) \right] \Big \} \nonumber \\
&+& \frac{g_Z^2}{64 \pi^2}\Big [ 4(c_W^2-s_W^2)^2 B_5 (p^2, m_{H^{\pm\pm}}, m_{H^{\pm\pm}})+ 
(c_W^2-s_W^2-c_{\beta_\pm}^2)^2 B_5 (p^2, m_{H^{\pm}}, m_{H^{\pm}}) \nonumber \\
&+&(c_W^2-s_W^2-s_{\beta_\pm}^2)^2 B_5 (p^2, m_{G^{\pm}}, m_{G^{\pm}}) +
2 s_{\beta_\pm}^2 c_{\beta_\pm}^2 B_5(p^2, m_{H^\pm}, m_{G^\pm}) \nonumber \\
 &+&(2c_\alpha c_{\beta_0} +s_\alpha s_{\beta_0})^2 B_5(p^2,m_H, m_A) 
+(2s_\alpha c_{\beta_0} -c_\alpha s_{\beta_0})^2 B_5(p^2,m_h, m_A) \nonumber \\
  &+&(s_\alpha c_{\beta_0} -2 c_\alpha s_{\beta_0})^2 B_5(p^2,m_H, m_{G^0})
+(c_\alpha c_{\beta_0} +2s_\alpha s_{\beta_0})^2 B_5(p^2,m_h, m_{G^0}) \Big ].
\end{eqnarray}
The photon-$Z$-boson mixing is calculated as:
\begin{eqnarray}
\Pi_{\rm HTM}^{Z \gamma}(p^2)&=&\frac{g^2 s_W}{16 \pi^2 c_W} \Big \{ \frac{g^2}{2} \sqrt{v_\phi^2+2 v_\Delta^2} \Big [ v_\phi s_W^2c_{\beta_\pm}+\sqrt{2} v_{\Delta} (1+s_W^2) s_{\beta_\pm} \Big ] B_0(p^2, m_W, m_W)\nonumber \\
& -&2 (c_W^2-s_W^2)B_5 (p^2, m_{H^{\pm\pm}}, m_{H^{\pm\pm}})
- \frac12 (c_{W}^2 -s_W^2- c_{\beta_\pm}^2) B_5(p^2,m_{H^\pm}, m_{H^\pm}) \nonumber \\
&- &\frac12 (c_{W}^2 -s_W^2- s_{\beta_\pm}^2) B_5(p^2,m_{W}, m_{W}),
\end{eqnarray}
where we used the short-hand notation for the Higgs  mixing angles $s(c)_\alpha \equiv \sin(\cos)\alpha$, $s(c)_{\beta_0} \equiv \sin(\cos){\beta_0}$, $s(c)_{\beta_\pm} \equiv \sin(\cos){\beta_\pm}$,  and for $s(c)_W \equiv \sin(\cos) \theta_W$. Here $m_{G^{\pm}}$ and $m_{G^0}$ are the  masses of the Nambu-Goldstone bosons $G^{\pm}$ and $G^0$, respectively, which in the 'Hooft-Feynman gauge, are the same as the corresponding gauge boson masses i.e. $m_{G^{\pm}}=m_W$ and $m_{G^0}=m_Z$. The $B_0-B_5$ functions are listed in \cite{Hagiwara:1994pw}.  

\begin{figure}[htbp]
\vskip -0.3in
\begin{center}$
    \begin{array}{cc}
    \includegraphics[width=3.6in,height=3.8in]{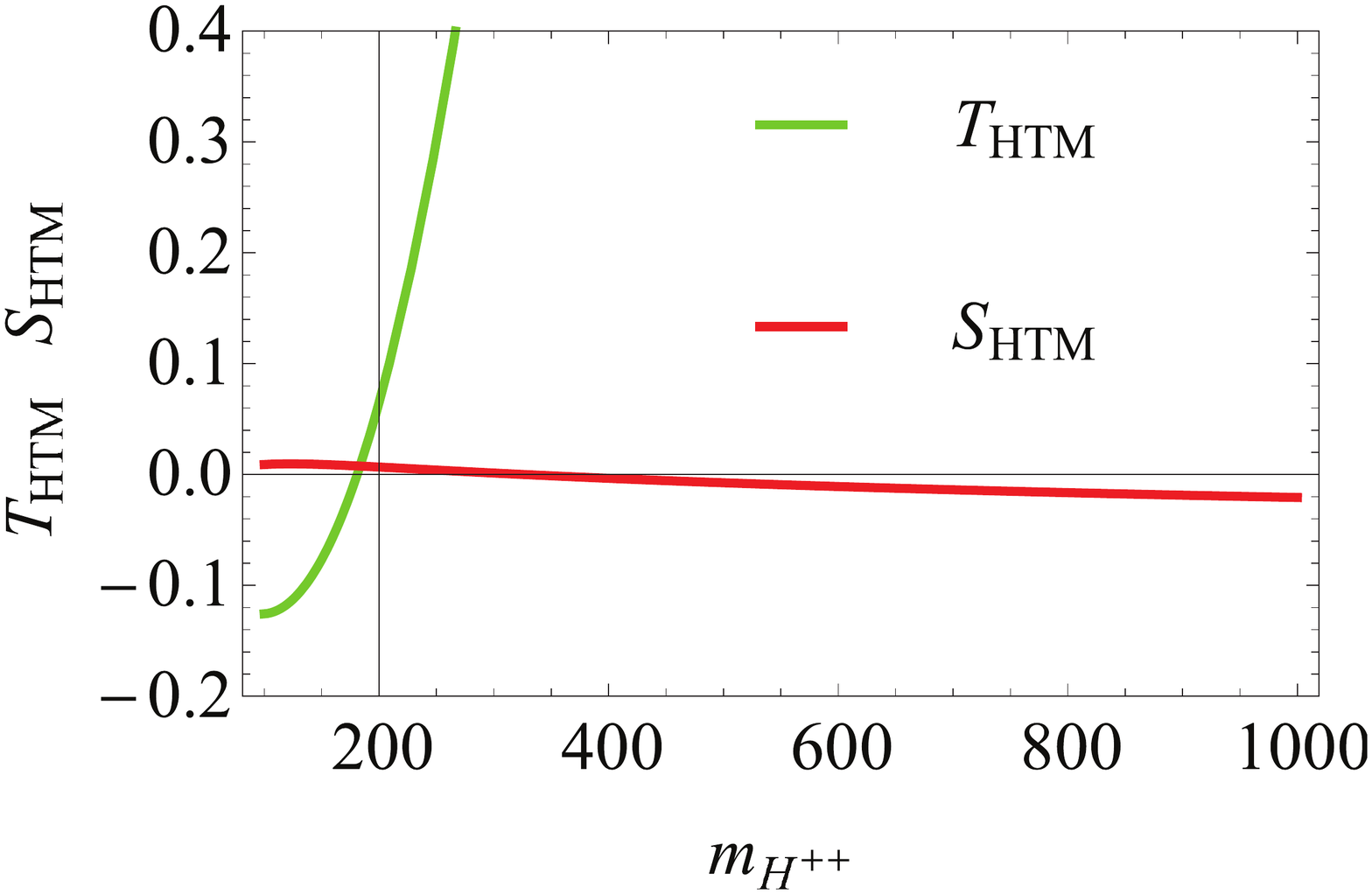}
&\hspace*{-1.5cm}
    \includegraphics[width=3.6in,height=3.8in]{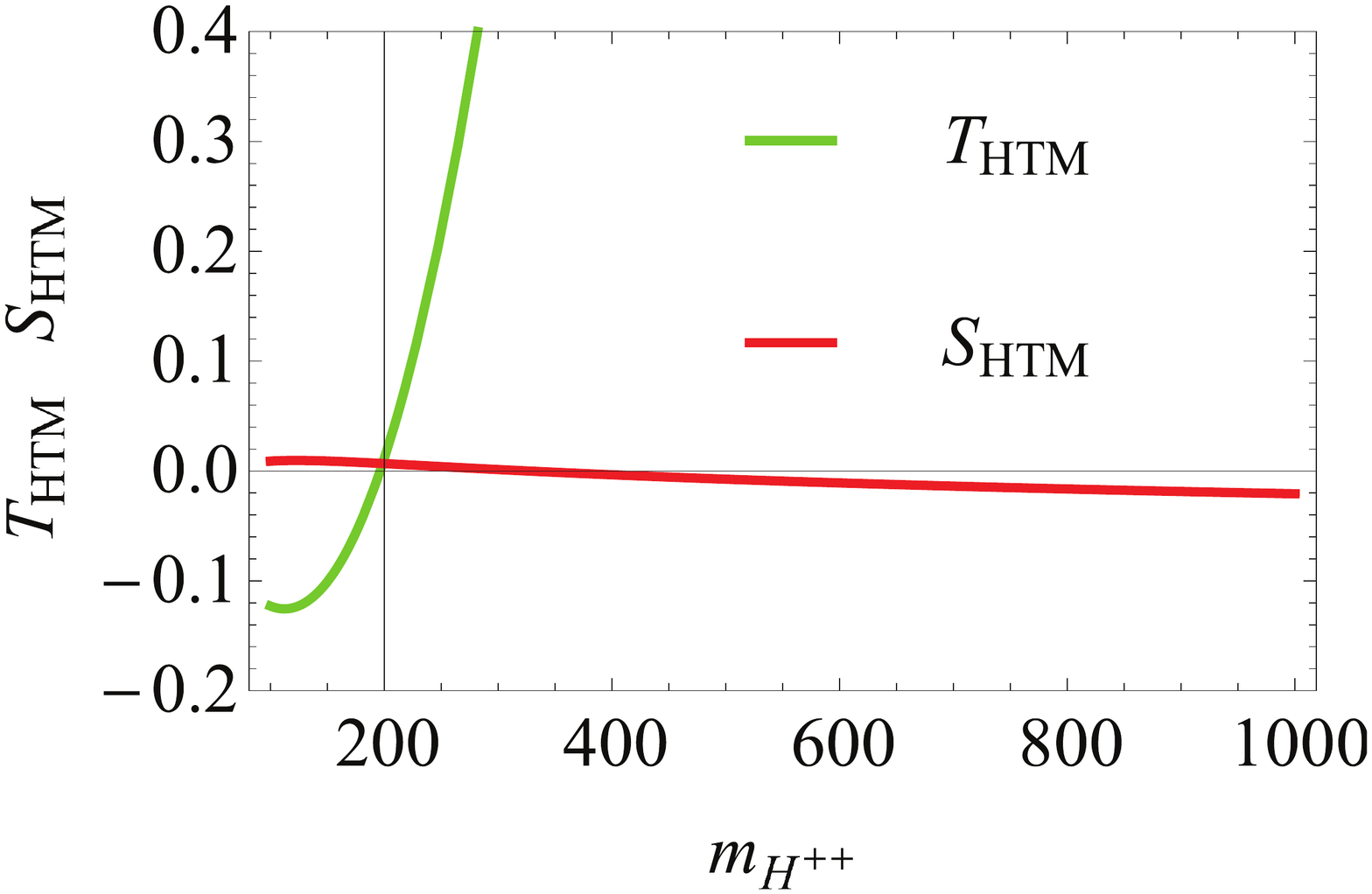}
        \end{array}$
\end{center}
\vskip -0.4in
     \caption{(color online). \sl\small The contribution to the $T$ and $S$ parameters in the HTM, as a function of the doubly charged Higgs mass, (left) for $\sin \alpha=0$, (right) for $\sin \alpha=1$. We take $v_\Delta=1$ GeV and indicate the allowed regions for $\Delta T$.}
\label{fig:TSinHTM}
\end{figure}

In  Fig. \ref{fig:TSinHTM} we show the dependence of the $T$ and $S$ parameters on the doubly charged Higgs mass, for $v_\Delta=1$ GeV, for the minimum mixing in the neutral sector, $\sin \alpha =0$, in the left panel, and maximum mixing, $\sin \alpha =1$,  in the right panel. The vertical axes are chosen to indicate the experimental limits. The figure shows that while the $S$ parameter agrees with experimental constraints over the whole parameter space, the $T$ parameter is very sensitive to the doubly doubly charged Higgs mass and,  if constrained to lie the allowed range,  an upper bound on $m_{H^{\pm \pm}}$ of $\sim 266$ GeV is required, for the case of no mixing in the neutral CP-even Higgs sector. The bound is slightly raised for $\sin \alpha=1$ (maximal mixing), but not significantly (upper bound on $m_{H^{\pm \pm}}$ of $\sim 280$ GeV). Varying the triplet VEV $v_\Delta $ does not affect the results significantly. There results agree with previous studies \cite{Kanemura:2012rs} and represent a potential problem for the HTM, as they are in apparent conflict with the experimental limits on the doubly charged mass, as summarized below. 

The mass of doubly charged Higgs boson $m_{H^{\pm\pm}}$ has been constrained by the Large Electron Positron Collider (LEP) \cite{OPAL}, the Hadron Electron Ring Accelerator (HERA) \cite{H1} and the Tevatron \cite{D0}. Some restrictions have been obtained independently of the decay modes of the boson. Particularly, if  $m_{H^{\pm\pm}}$ is less than  half of the $Z$ boson mass, the new decay mode $Z \rightarrow H^{\pm\pm}H^{\mp\mp}$ will open.  From the precise measurement of total decay width of the Z boson $\Gamma_Z^{NP} < 3 ~ \textrm{MeV}~ (95\% \textrm{C.L.})$ \cite{Beringer:1900zz}, and the partial decay width into a doubly charged boson pair, a
 lower mass bound  $m_{H^{\pm\pm}} > 42.9$ GeV at $95\%$ C.L. can be obtained.

The most up-to-date mass bounds have been obtained  through the direct searches at the LHC.  The ATLAS Collaboration has looked for doubly charged Higgs bosons via pair production in the same sign di-lepton final states. Based on the data sample corresponding to an integrated luminosity of 4.7 $\text{fb}^{-1}$ at $\sqrt{s} = 7$ TeV, the masses below 409 GeV, 375 GeV and 398 GeV have been excluded respectively for $e^{\pm}e^{\pm}$, $e^{\pm}\mu^{\pm}$ and $\mu^{\pm}\mu^{\pm}$ by assuming a branch ratio of $100\%$ for each final state \cite{ATLAS:2012ai}. 
The CMS Collaboration also considered the associated production
$pp \rightarrow H^{\pm\pm}H^{\mp}$, in which the masses of $H^{\pm\pm}$ and $H^{\mp}$ are assumed to be degenerate. Using three or more isolated charged lepton final states, the lower limit on $m_{H^{\pm\pm}}$  was found to be between 204 and 459 GeV in the 100\% branching fraction scenarios, and between 383 and 408  GeV for the type II see-saw scenarios  \cite{Chatrchyan:2012ya}. These limits raise doubts about the existence of a light doubly charged Higgs boson.

However, the constraints have been questioned by several authors. Other decay modes for $H^{\pm\pm}$ such as those into $W$-pairs become dominant under some conditions, namely $v_\Delta \gsim 10^{-4}$ GeV \cite{Chiang:2012dk}. Using the ATLAS result (with $4.7 \text{fb}^{-1}$ integrated luminosity at $\sqrt{s} = 7$ TeV) from the search by the lepton-pair production, these authors obtain a lower limit for the doubly charged boson mass of 60 GeV at the $95\%$ C.L., re-evaluated to be 85 GeV for an  integrated luminosity of $20 \text{fb}^{-1}$. 

Still, the window for observing a light (left-handed) doubly charged Higgs boson is fairly narrow, and it would be desirable that a viable model should be able to accommodate heavier masses for these bosons. In the next section, we shall see that the upper bounds  on doubly charged masses from precision electroweak constraints are raised by introducing vector quarks.  

\subsection{Vector-Like  Quark contributions to the $S$ and $T$ parameters}
\label{subsec:TSVQ}
The oblique correction parameter $S$ for vector-like quarks is\cite{Lavoura:1992np}:
\begin{eqnarray} 
S &=& \frac{N_c}{2 \pi} \left \{ \sum_\alpha \sum_i \left [ (|V_{\alpha i}^L|^2+ |V_{\alpha i}^R|^2 )\Psi_{+}(y_\alpha , y_i) +2 {\rm Re} (V_{\alpha i}^L V_{\alpha i}^{R\,\star}) \Psi_{-}(y_\alpha , y_i) \right  ]\right. \nonumber \\
&-& \left. \sum_{\beta<\alpha} \left [ (|U_{\alpha \beta}^L|^2+ |U_{\alpha \beta}^R|^2 )\chi_{+}(y_\alpha , y_\beta) +2 {\rm Re} (U_{\alpha \beta}^L U_{\alpha \beta}^{R\,\star}) \chi_{-}(y_\alpha , y_\beta) \right ] \right. \nonumber \\
 &-& \left.\sum_{j<i} \left [ (|D_{ij}^L|^2+ |D_{ij}^R|^2 )\chi_{+}(y_i , y_j) +2 {\rm Re} (D_{ij}^L D_{ij}^{R\,\star}) \chi_{-}(y_i , y_j) \right ] \right \}
\end{eqnarray}
where the functions $\chi_{+(-)}$ are defined as
\begin{eqnarray}
\chi_{+} (y_1, y_2) &\equiv& \frac{y_1+y_2}{2}-\frac{(y_1-y_2)^2}{3}+ \Big [\frac{(y_1-y_2)^3}{6}-\frac{1}{2} \frac{y_1^2+y_2^2}{y_1-y_2}\Big ] \ln\frac{y_1}{y_2}+\frac{y_1-1}{6}f(y_1,y_1) \nonumber \\
&+&\frac{y_2-1}{6}f(y_2,y_2)+\Big [\frac{1}{3}-\frac{y_1+y_2}{6}-\frac{(y_1-y_2)^2}{6}\Big ]f(y_1,y_2), \nonumber \\
\chi_{-} (y_1, y_2) &\equiv& -\sqrt{y_1 y_2} \Big [2+(y_1- y_2-\frac{y_1+ y_2}{y_1-y_2}) \ln\frac{y_1}{y_2}+\frac{f(y_1,y_1)+f(y_2,y_2)}{2}-f(y_1,y_2)\Big ],\nonumber \\
\end{eqnarray}
and the function $f$ is:
\begin{eqnarray}
f(y_1,y_2) &\equiv& \left\{
\begin{array}{ll}  \displaystyle
-2\sqrt{\Delta}\left(\arctan\frac{y_1-y_2+1}{\sqrt{\Delta}}-\arctan\frac{y_1-y_2-1}{\sqrt{\Delta}}\right ) & \Delta>0 \\
\displaystyle
0 & \Delta=0 \\
\displaystyle \sqrt{-\Delta}\ln\frac{y_1+y_2-1+\sqrt{-\Delta}}{y_1+y_2-1-\sqrt{-\Delta}} \hspace{0.5cm} & \Delta<0 \, ,
\end{array} \right. 
\label{eq:fy} 
\end{eqnarray}
where 
$
\Delta= -1-y_1^2-y_2^2+2y_1+2y_2+2y_1y_2.$
 The functions $\Psi_{+}$ and $\Psi_{-}$ are defined by
\begin{eqnarray}
\Psi_{+} (y_\alpha, y_i) &\equiv& \frac{22y_\alpha+14y_i}{9}-\frac{1}{9}\ln\frac{y_\alpha}{y_i}+\frac{11y_\alpha+1}{18}f(y_\alpha,y_\alpha)+\frac{7y_i-1}{18}f(y_i,y_i),\nonumber \\ 
\Psi_{-} (y_\alpha, y_i) &\equiv& -\sqrt{y_\alpha y_i} \Big [4+\frac{f(y_\alpha,y_\alpha)+f(y_i,y_i)}{2}\Big ].
\end{eqnarray}

\begin{figure}[htbp]
\vskip -0.3in
\begin{center}$
    \begin{array}{cc}
\hspace*{-1.0cm}
    \includegraphics[width=3.3in,height=3.6in]{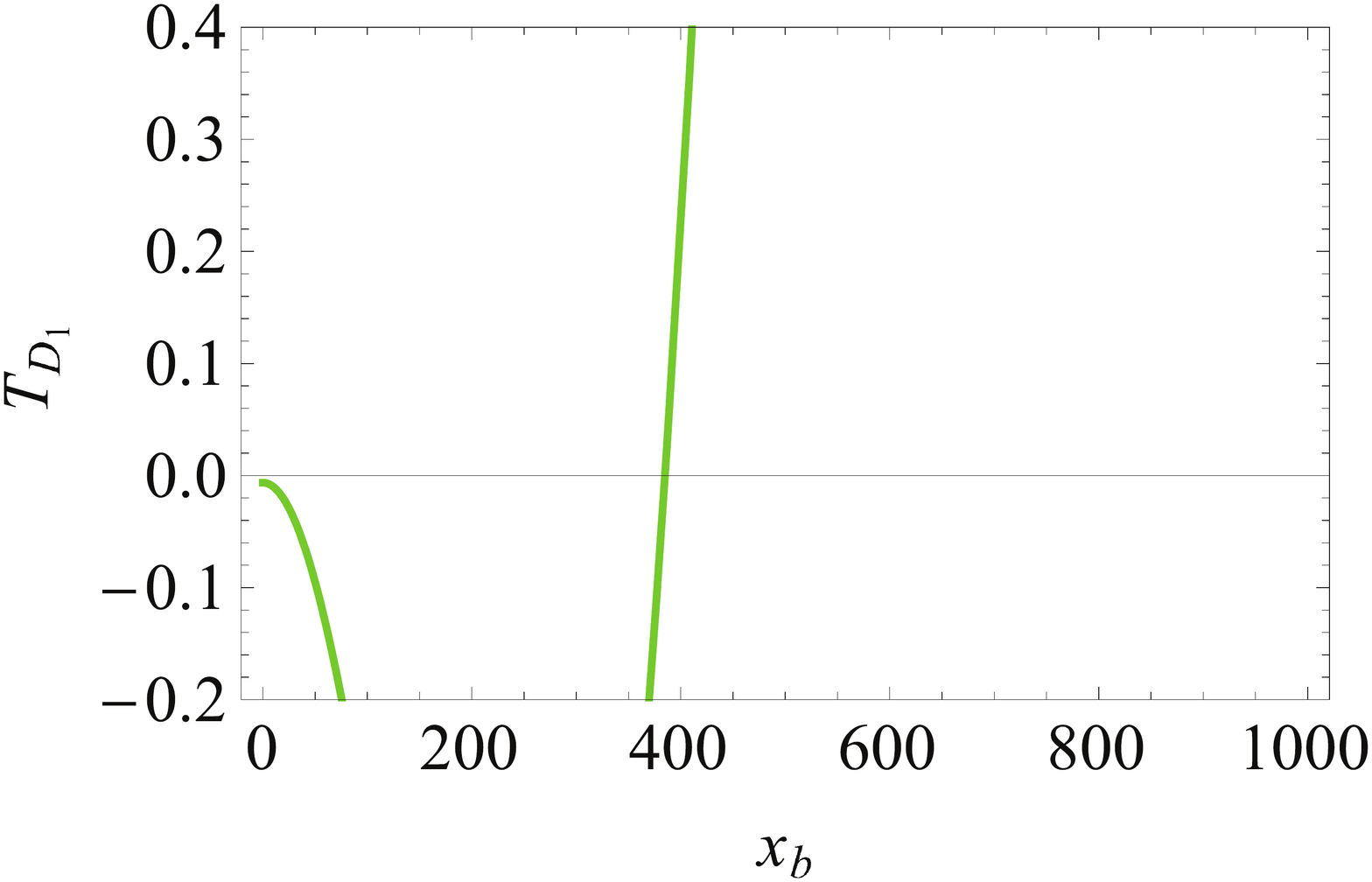}
&
    \includegraphics[width=3.3in,height=3.6in]{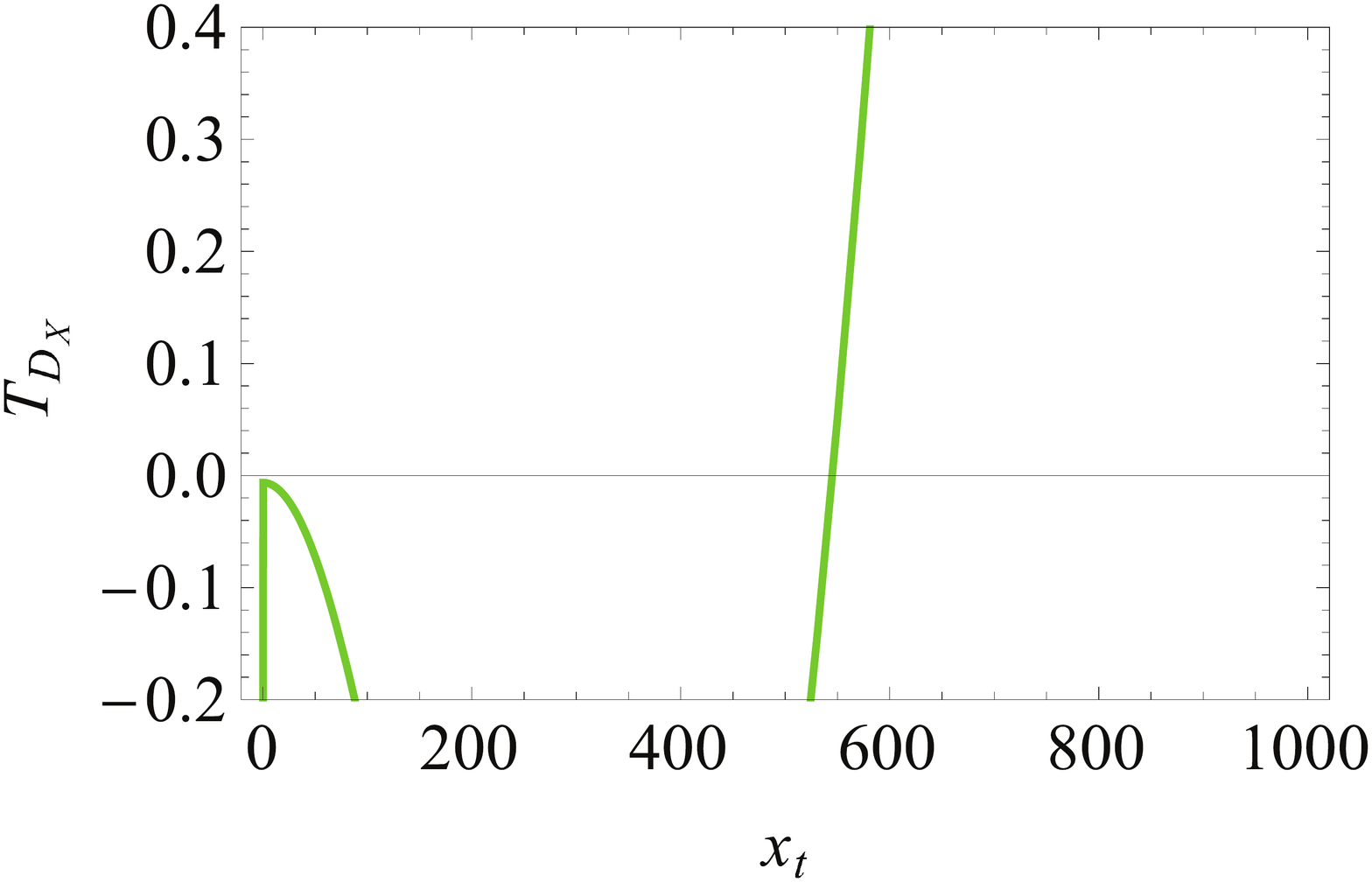}
       \end{array}$
\end{center}
\vskip -0.4in
 \caption{ (color online). \sl\small The allowed variation of the $T$ parameter with $x_{t(b)}$, as defined in the text, for models ${\cal D}_1$ (left panel), ${\cal D}_X$ ( right panel). We chose $M=350$ GeV for both plots.} \label{fig:Tfctxt}
\end{figure}

The oblique correction parameter $T$ for vector-like quarks is:
\begin{eqnarray} 
T &=& \frac{N_c}{16 \pi s^2_W c^2_W} \left \{ \sum_\alpha \sum_i \left [ (|V_{\alpha i}^L|^2+ |V_{\alpha i}^R|^2 )\theta_{+}(y_\alpha , y_i) +2 {\rm Re} (V_{\alpha i}^L V_{\alpha i}^{R\,\star}) \theta_{-}(y_\alpha , y_i) \right  ]\right. \nonumber \\
&-& \left. \sum_{\beta<\alpha} \left [ (|U_{\alpha \beta}^L|^2+ |U_{\alpha \beta}^R|^2 )\theta_{+}(y_\alpha , y_\beta) +2 {\rm Re} (U_{\alpha \beta}^L U_{\alpha \beta}^{R\,\star}) \theta_{-}(y_\alpha , y_\beta) \right ] \right. \nonumber \\
 &-& \left.\sum_{j<i} \left [ (|D_{ij}^L|^2+ |D_{ij}^R|^2 )\theta_{+}(y_i , y_j) +2 {\rm Re} (D_{ij}^L D_{ij}^{R\,\star}) \theta_{-}(y_i , y_j) \right ] \right \}, 
\end{eqnarray}
where $V_{\alpha i}^{L,R}$, $U_{\alpha \beta}^{L,R}$ and $D_{ij}^{L,R}$ are listed in Appendix.  We adopted the convention of using Greek letters to denote up-type quarks and Latin ones to denote down-type quarks. Here $N_c=3$ is the number of colours, and the functions $\theta_{+(-)}$ are defined as
\begin{eqnarray}
\theta_{+} (y_1, y_2) &\equiv& y_1+y_2-\frac{2y_1 y_2}{y_1-y_2} \ln\frac{y_1}{y_2},\nonumber \\ 
\theta_{-} (y_1, y_2) &\equiv& 2 \sqrt{y_1 y_2} \left (\frac{y_1+ y_2}{y_1-y_2} \ln\frac{y_1}{y_2}-2\right ),
\end{eqnarray}
where $\displaystyle y_i=\frac{m_i^2}{m_Z^2}$ \cite{Lavoura:1992np}. As in the HTM without vector-like quarks, the $S$-parameter does not impose any restrictions on the parameter space of the model. We concentrate on the $T$ parameter. 
As explicit expressions exist for the $T$ parameter in some models \cite{Cai:2012ji},  we do not include them all. We are interested in the case in which the contributions from vector-like quarks are of opposite signs to those from the extra states in the HTM, and thus allow to relax the severe constraint on the doubly charged Higgs mass discussed in the previous section.  In Fig. \ref{fig:Tfctxt}, we show the contribution to the $T$ parameter in two of the models, ${\cal D}_1$, ${\cal D}_X$. We have chosen these models since these are the only ones which  yield  contributions to the $T$ parameter which can be negative, interfering destructively with those coming from the particle content of the HTM. As shown in Fig.  \ref{fig:Tfctxt},  the $T$ parameter in these models is negative in a small region, restricting  the upper bound on $m_{H^{\pm \pm}}$ to  $\sim 400$ GeV in the  ${\cal D}_X$  and ${\cal D}_1$ models.  The rest of the models from Table \ref{tab:VQrepresentations} not shown in Fig. \ref{fig:Tfctxt} give always a positive contribution to the $T$ parameter and thus, when added to the HTM contribution, the restrictions on the doubly charged Higgs mass worsen.

\subsection{Restrictions on doubly charged Higgs boson and vector-like quarks masses} 
\label {subsec:mHpp}
We investigate further the ${\cal D}_1$ and ${\cal D}_X$ models, where the negative contributions to the $T$ parameter are significant. For the specific models under study, we give explicit expressions for the $T$ parameter\footnote{Explicit expressions for model ${\cal D}_X$ appear in \cite{Cai:2012ji}, but we include it here,  and add expressions for model ${\cal D}_1$ for completeness.}:
\begin{eqnarray}
\Delta T_{{\cal D}_1}&=&\frac{3}{16 \pi s^2_W c^2_W } \left[s_L^{d\,2} \theta_{+}(y_t, y_B)-s_L^{d\,2} \theta_{+}(y_t,y_b)- c_L^{d\,2} s_L^{d\,2} \theta_{+}(y_b, y_B) \right], \nonumber \\
\Delta T_{{\cal D}_X}&=&\frac{3}{16 \pi s^2_W c^2_W} \Big [ s_L^{u\,2} \theta_{+}(y_T,y_b)-s_L^{u\,2} \theta_{+}(y_t, y_b) 
+\left( s_L^{u\,2}+s_R^{u\,2} \right )\theta_{+}(y_t, y_X)\nonumber \\
&+& \left( c_L^{u\,2}+c_R^{u\,2} \right )\theta_{+}(y_T, y_X) 
+ 2 s_L^{u} s_R^{u} \theta_{-}(y_t, y_X) + 2 c_L^{u} c_R^{u} \theta_{-}(y_T, y_X) \nonumber \\
&-& \left (4 c_L^{u\,2} s_L^{u\,2} + c_R^{u\,2} s_R^{u\,2} \right ) \theta_{+}(y_t, y_T) 
- \left (4 c_L^{u} s_L^{u} c_R^{u} s_R^{u} \right )\theta_{-}(y_t, y_T)
\Big ]. 
\end{eqnarray}
For a given physics model, the predictions for the  $T$  parameter consist of the sum of the vector-quark contributions and the non-vanishing SM reminders, when the Higgs mass ($m_h$) and top mass ($m_t$) differ from those used for the SM reference. The dependence of $T$ on the latter two parameters is then approximated by the one-loop terms
\begin{equation}
\Delta T_{h,\,t} \sim -\frac{3}{16 \pi c^2_W} \ln \frac{m_h^2}{m_{h, \rm ref}^2} +\frac{3}{16 \pi s^2_W c^2_W} \ln \frac{m_t^2-m^2_{t, \rm ref}}{m_Z^2}.
\end{equation}
The $m_t$ dependence is often neglected \cite{Baak:2011ze}. Assuming the Higgs mass is 125 GeV and its reference value $m_{h, \rm ref}=120$ GeV \cite{Cai:2012ji}, we added the two sources (vector quark contributions and the correction coming from the Higgs mass deviation from its reference value) to the $T$ parameter in the HTM.

\begin{figure}[htbp]
\begin{center}$
    \begin{array}{cc}
\hspace*{-1.7cm}
    \includegraphics[width=3.8in,height=3.2in]{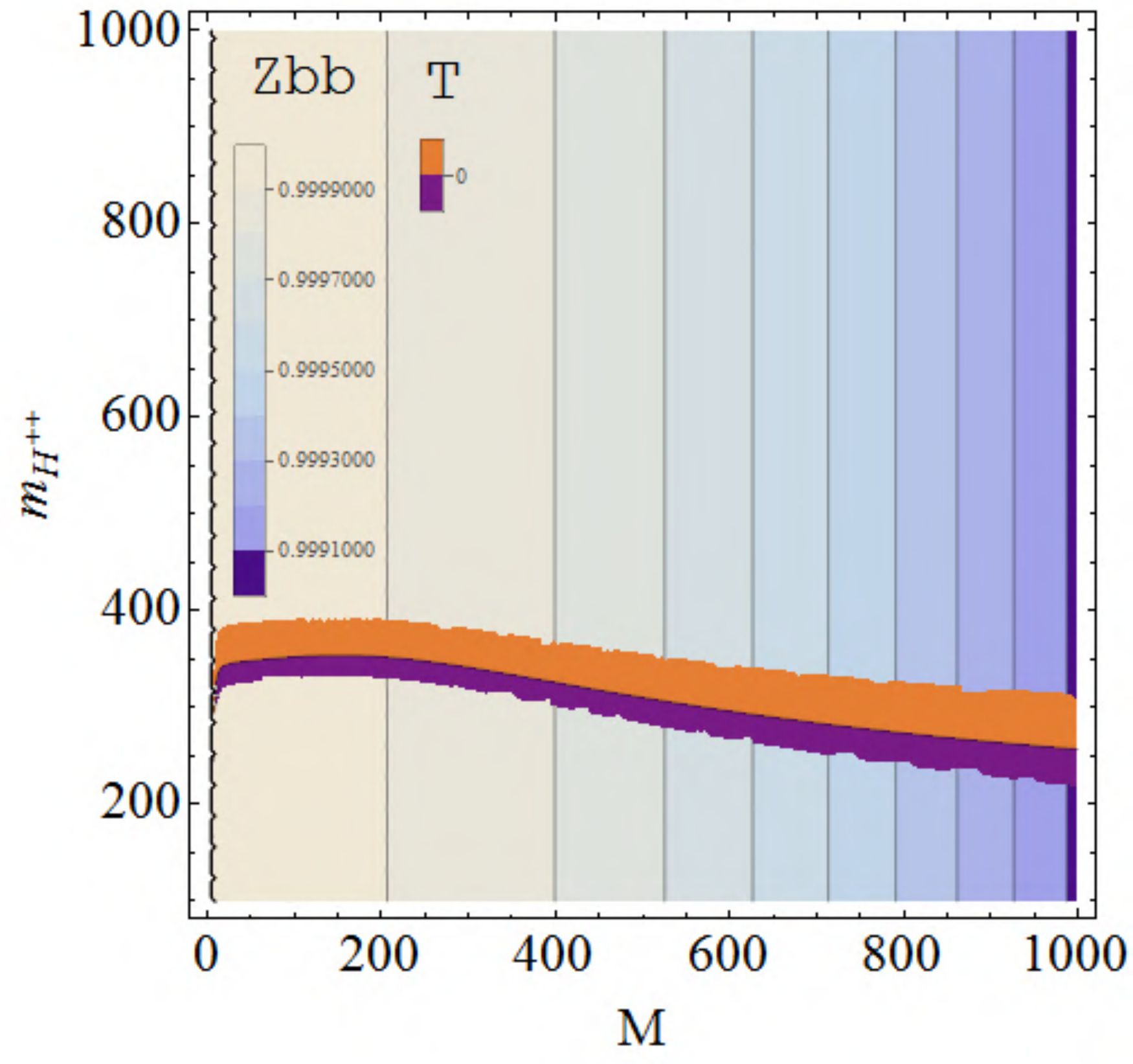}
&
    \includegraphics[width=3.8in,height=3.2in]{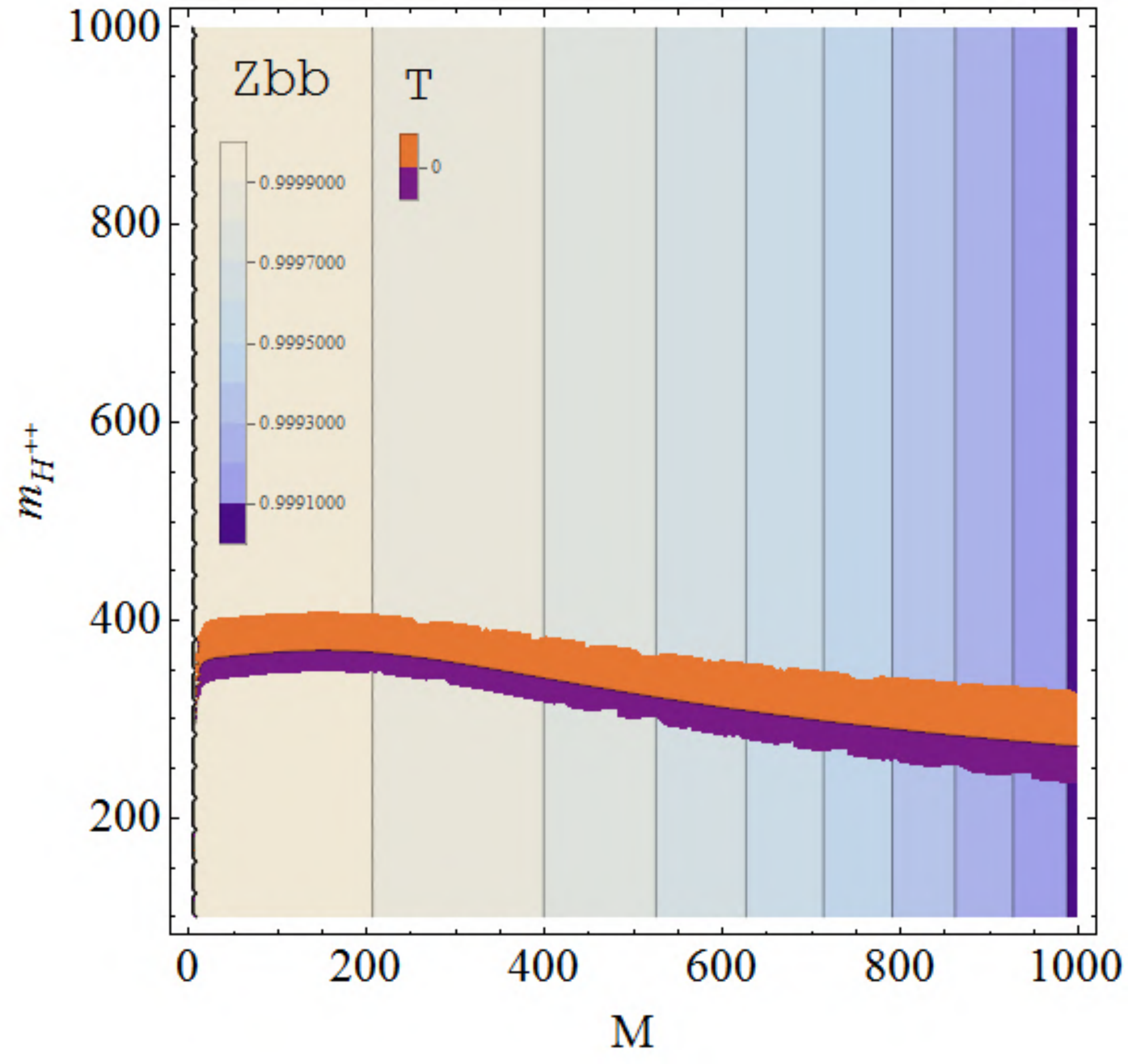}\\
    \hspace*{-1.7cm}
    \includegraphics[width=3.8in,height=3.2in]{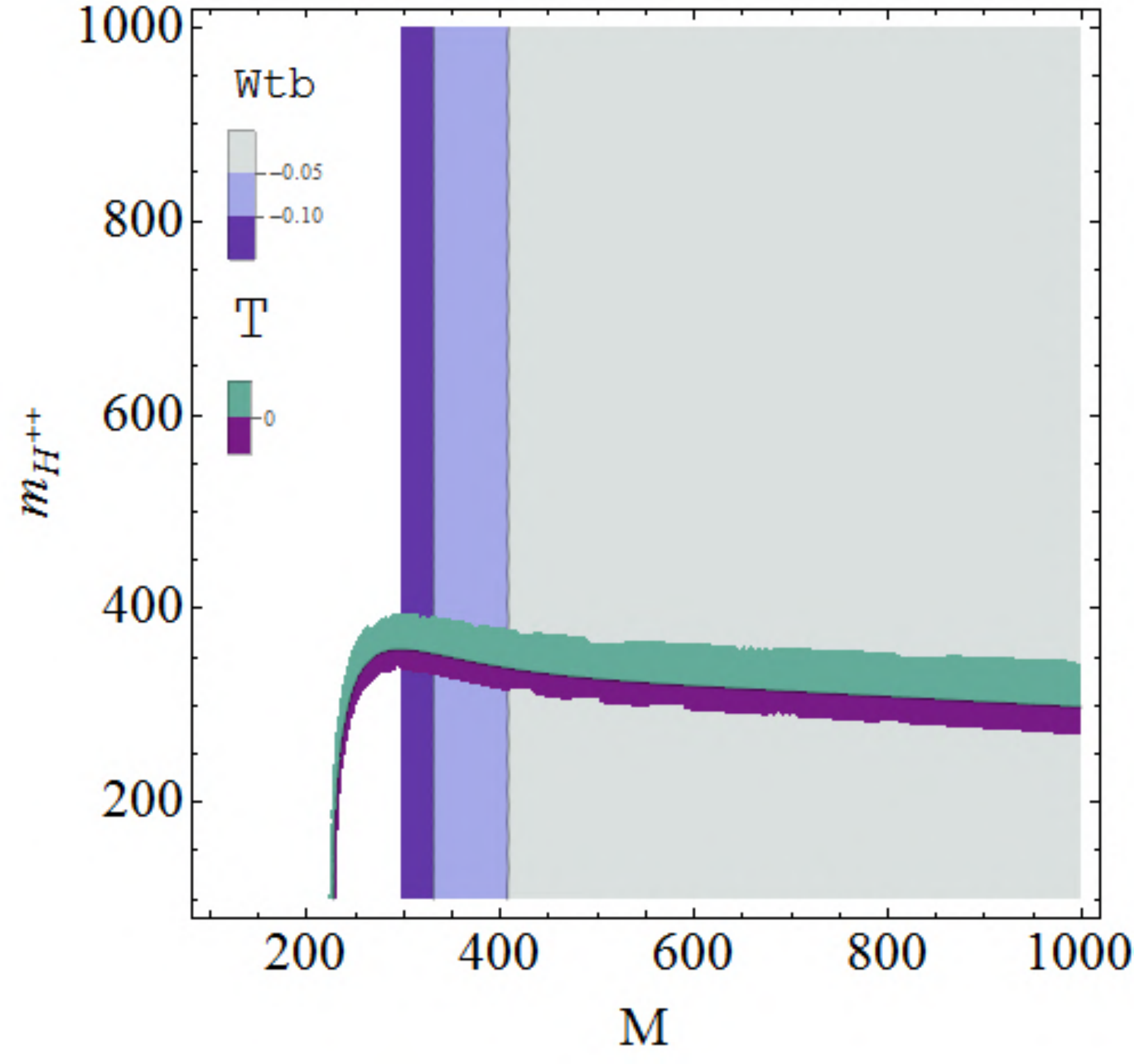}
&
    \includegraphics[width=3.8in,height=3.2in]{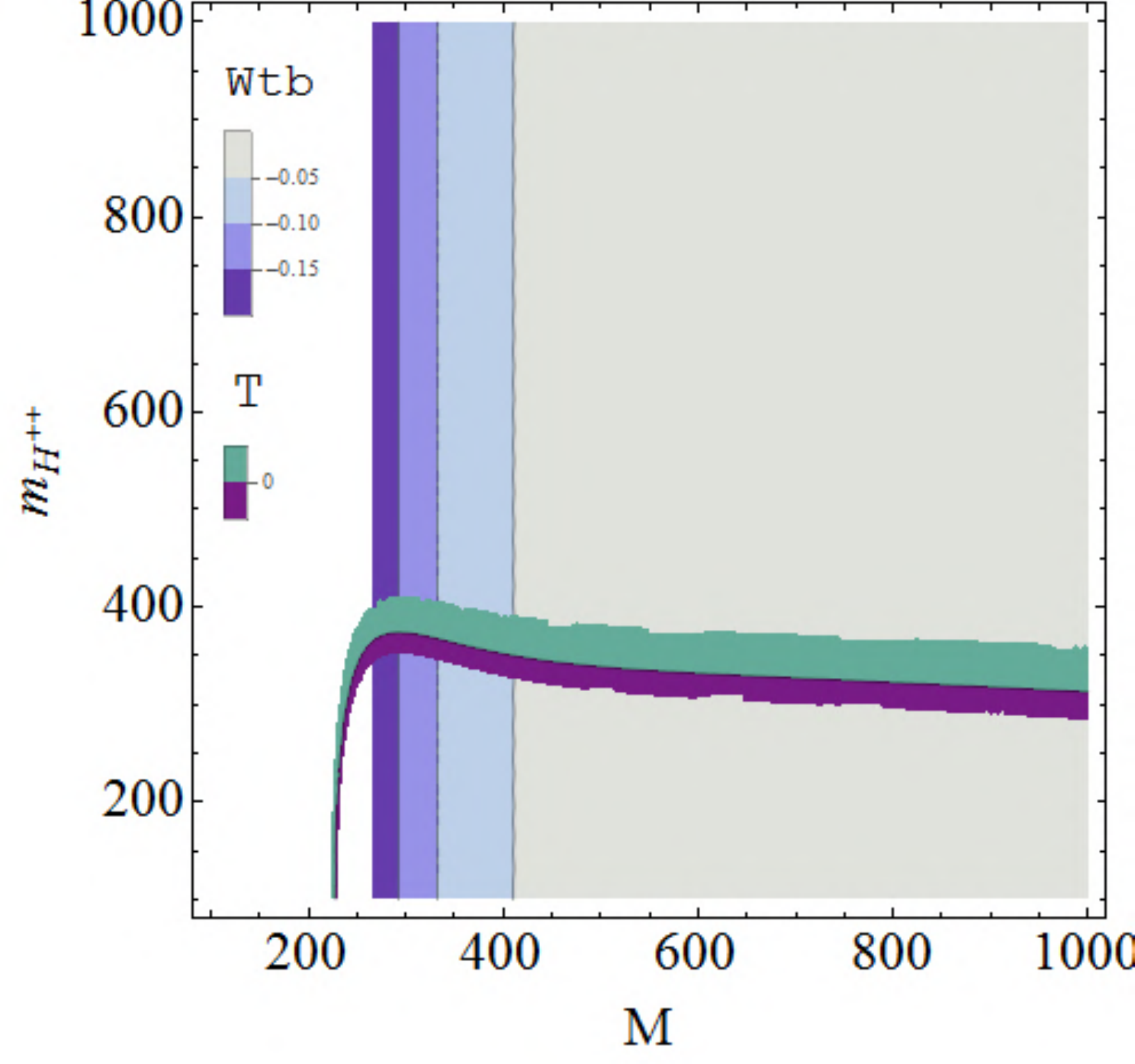}\\
        \end{array}$
\end{center}
\vskip -0.1in
     \caption{(color online). \sl\small Contour graphs showing the contribution to the $T$  parameter in the HTM with vector quarks, as functions of the doubly charged Higgs mass $m_{H^{\pm\pm}}$ and the vector quark mass $M$, for fixed values of $x_{b(t)}$. We show (upper left panel) the ${\cal D}_1$ model with $x_b=230$ GeV, $\sin \alpha=0$, (upper right panel) the ${\cal D}_1$ model with $x_b=230$ GeV, $\sin \alpha=1$, (lower left panel) the ${\cal D}_X$ model with $x_t=360$ GeV, $\sin \alpha=0$, (lower right panel) the ${\cal D}_X$ model with $x_t=370$ GeV, $\sin \alpha=1$.}
\label{fig:xtbTinD1DXHTM}
\end{figure}

Motivated by $T$ parameter contributions to ${\cal D}_1$ and ${\cal D}_X$ models from vector-like quarks which are of opposite sign from the contributions in the HTM, we proceed to analyze restrictions on the doubly charged Higgs mass when we add the singlet  $B$ vector-like quark in ${\cal D}_1$ scenario or the singlet $T$ vector-like quark in the non-SM vector-like doublet in the ${\cal D}_X$ scenario, to the particle representation of the Higgs Triplet Model. In Fig. \ref{fig:xtbTinD1DXHTM} we show the effects on the $T$ parameter as a contour in an $m_{H^{\pm \pm}}-M$ plane. The upper panels correspond to the ${\cal D}_1$ model, the lower to the ${\cal D}_X$ model. The left (right) panels correspond to no mixing (maximal mixing) in the neutral Higgs sector,  {\it i. e.}, between $h$ and $H$. In both models, it is clear that the presence of the mixing relaxes  the constraints  on the doubly charged mass from restrictions on the $T$-parameter, though generally by less than 10\%. In the ${\cal D}_1$ model, the maximum doubly charged mass value allowed is $m_{H^{\pm\pm}}=392$ GeV, reached for $x_b=230$ GeV for $\sin \alpha=0$, while for $\sin \alpha=1$, the maximum doubly charged mass value allowed is $m_{H^{\pm\pm}}=410$ GeV, for $x_b=230$ GeV. In the plots for the ${\cal D}_1$ model we include restrictions from $Z b {\bar b}$ decay,  while in the plots for the ${\cal D}_X$ model we include restrictions from $Wtb$ vertex. 

\begin{figure}[htbp]
\begin{center}$
    \begin{array}{cc}
\hspace*{-1.7cm}
    \includegraphics[width=3.8in,height=3.2in]{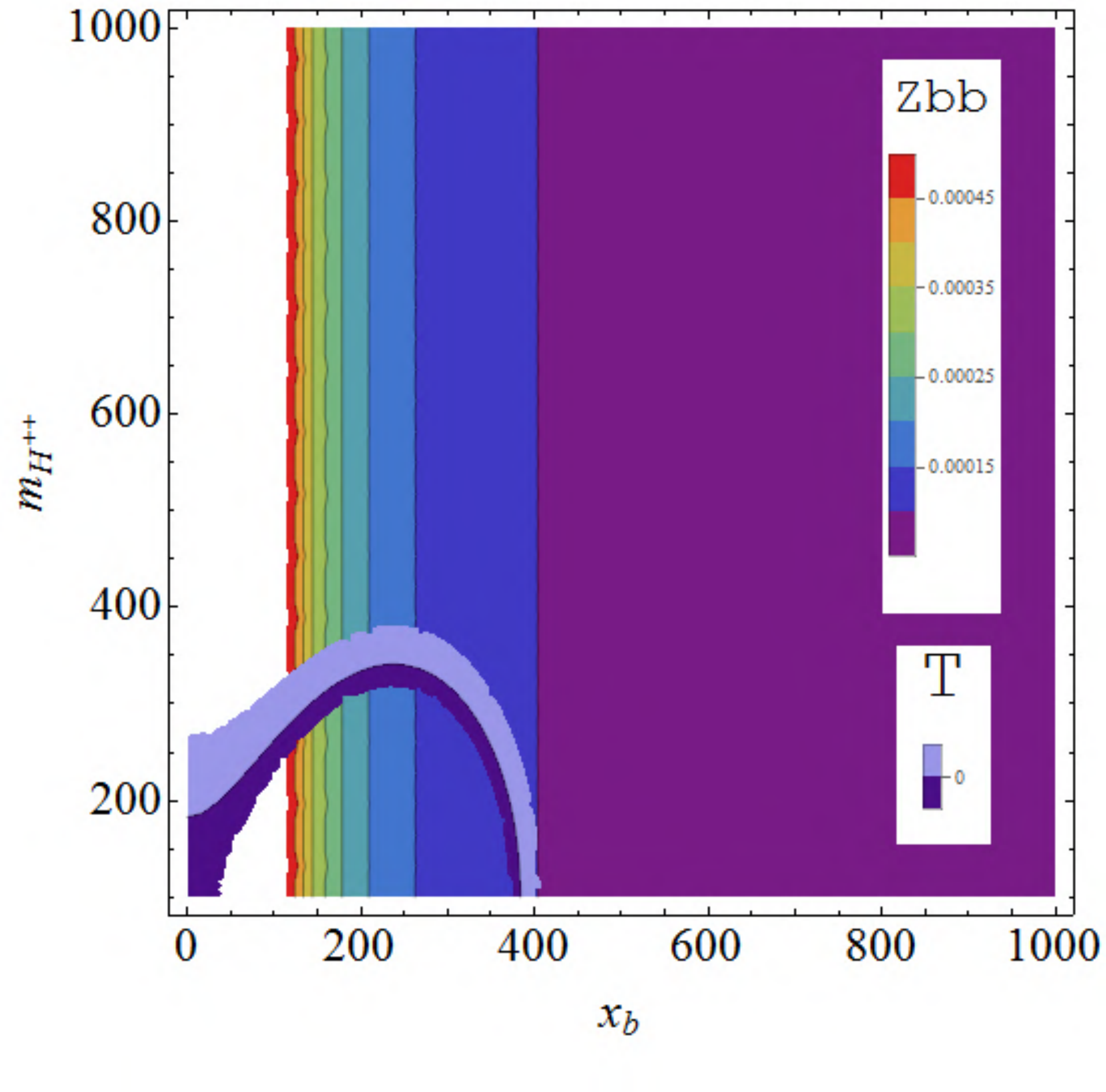}
&
    \includegraphics[width=3.8in,height=3.2in]{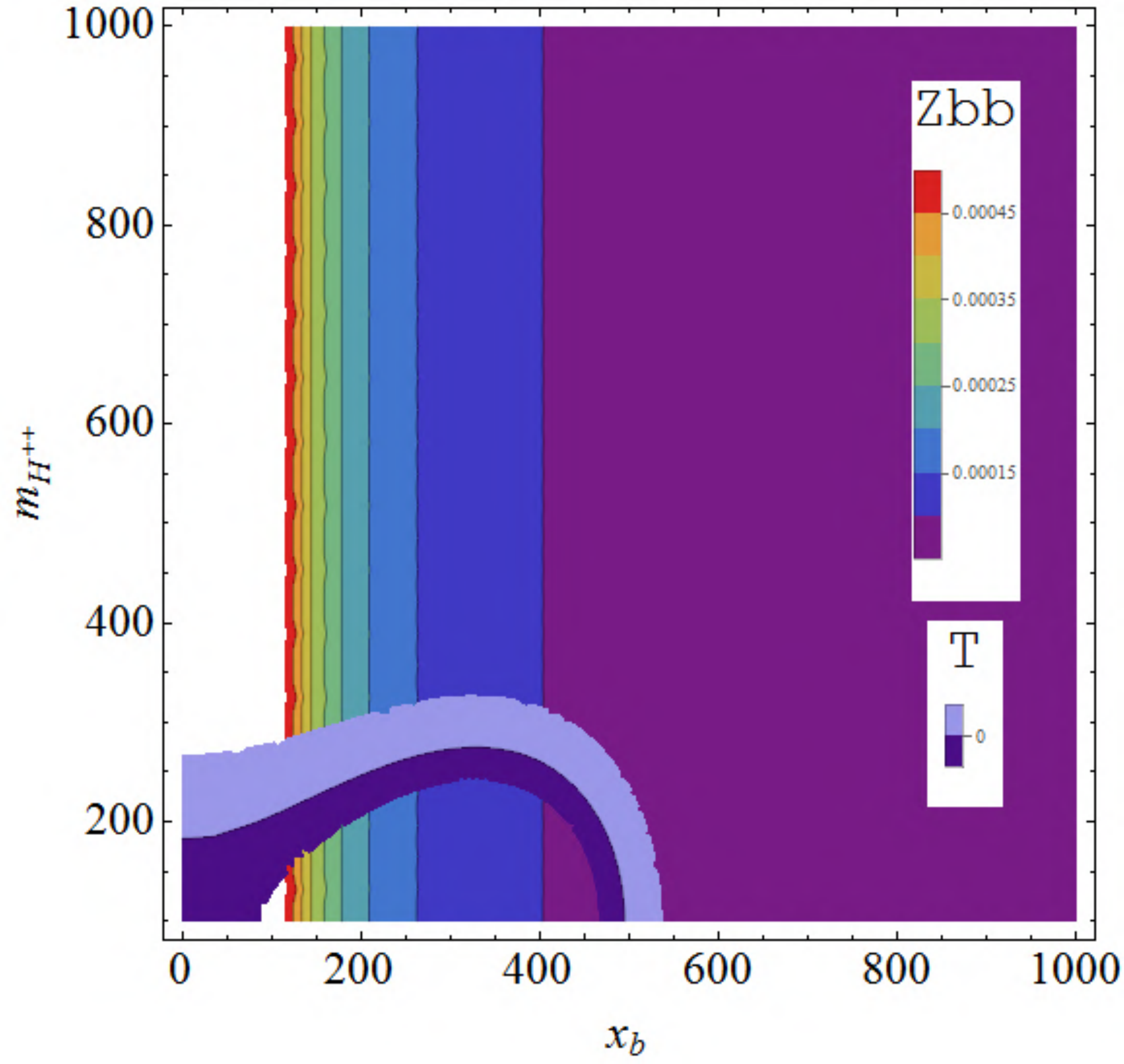}\\
    \hspace*{-1.7cm}
    \includegraphics[width=3.8in,height=3.2in]{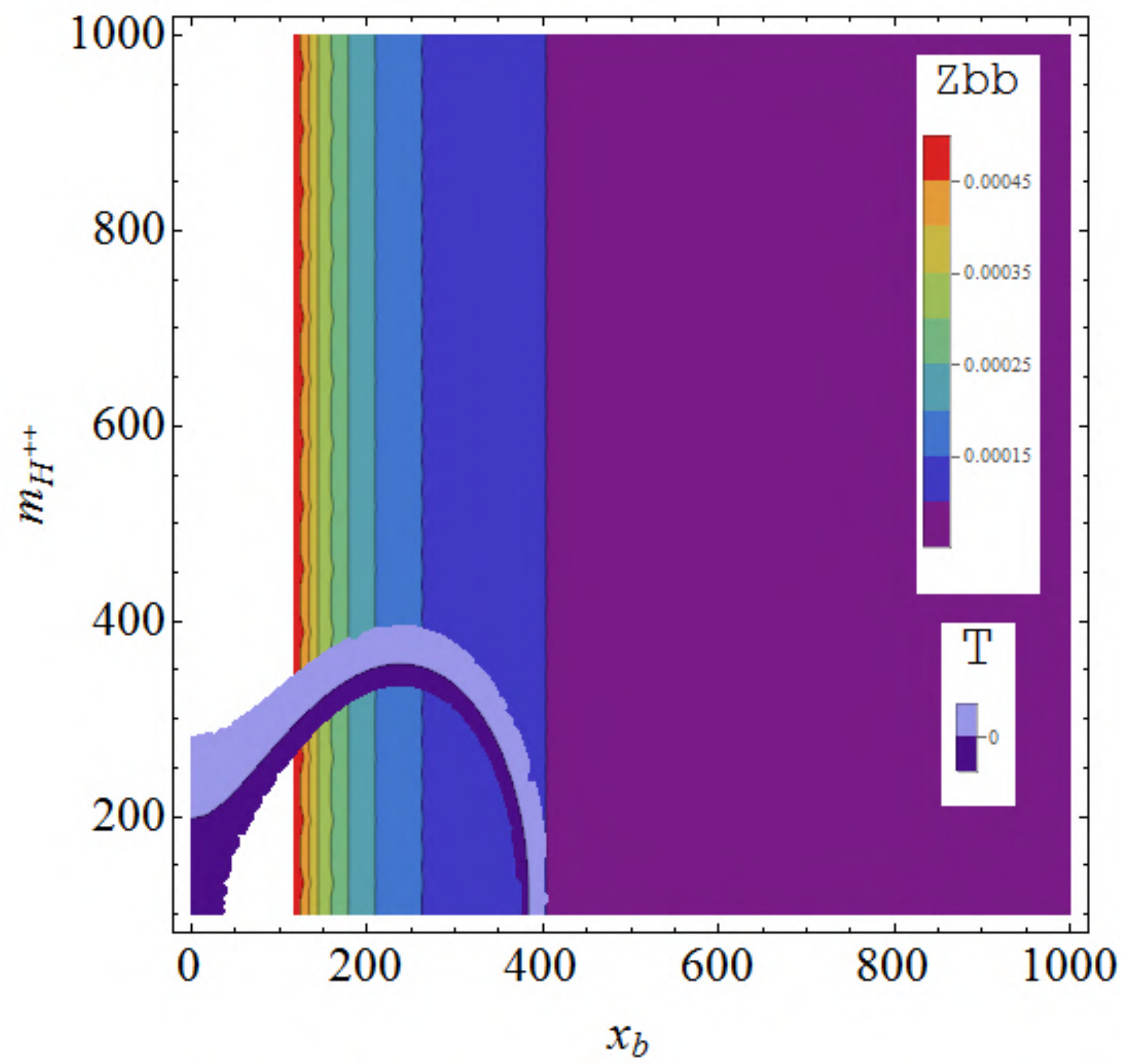}
&
    \includegraphics[width=3.8in,height=3.2in]{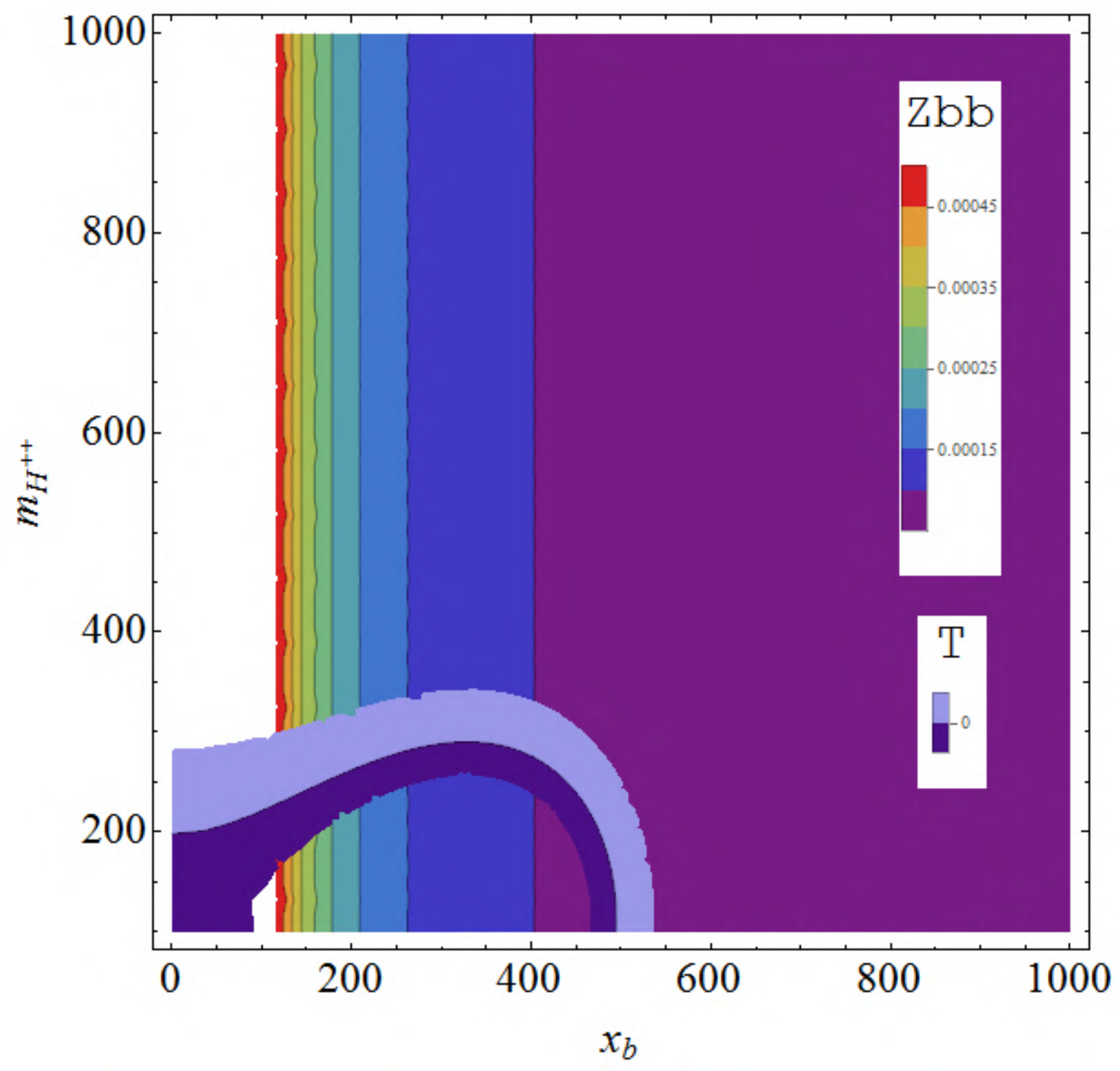}\\
        \end{array}$
\end{center}
\vskip -0.1in
     \caption{(color online). \sl\small Contour graphs showing the contribution to the $T$  parameter in the HTM with vector quarks, scenario ${\cal D}_1$, as functions of the doubly charged Higgs mass $m_{H^{\pm\pm}}$ and $x_b$, for fixed values of the vector quark mass $M$. We choose (upper left panel)  $M=305$ GeV, $\sin \alpha=0$, (upper right panel) $M=1000$ GeV, $\sin \alpha=0$, (lower left panel) $M=305$ GeV, $\sin \alpha=1$, (lower right panel) $M=1000$ GeV, $\sin \alpha=1$.}
\label{fig:MTinD1HTM}
\end{figure}

In the ${\cal D}_X$ scenario, the maximum doubly charged mass value allowed is $m_{H^{\pm\pm}}=397$ GeV, when $x_b=360$ GeV for $\sin \alpha=0$, while   for $\sin \alpha=1$, the maximum doubly charged mass value allowed is $m_{H^{\pm\pm}}=412$ GeV, for $x_b=370$ GeV.  The contour plots indicate the values for the $T$ parameter, as shown in the figure inserts. For all plots, we selected the particular value for $x_{t(b)}$ to correspond to the largest upper limit for doubly charged Higgs mass, as seen in Fig. \ref{fig:MTinD1HTM}.  

We note that, in the ${\cal D}_1$ model, for the chosen values of $x_{t(b)}$, the mass range for the vector-like quarks is not restricted by either the $T$ parameter or by $Z b {\bar b}$, while in the ${\cal D}_X$ model $Wtb$ constrains vector-like quarks to be $M \ge 300$ GeV.

We are interested in restriction on vector-like quark parameters $M$ and $x_{b(t)}$, which have further implications for doubly charged Higgs boson masses, as we shall discuss.


\begin{figure}[htbp]
\begin{center}$
    \begin{array}{cc}
\hspace*{-1.7cm}
    \includegraphics[width=3.8in,height=3.2in]{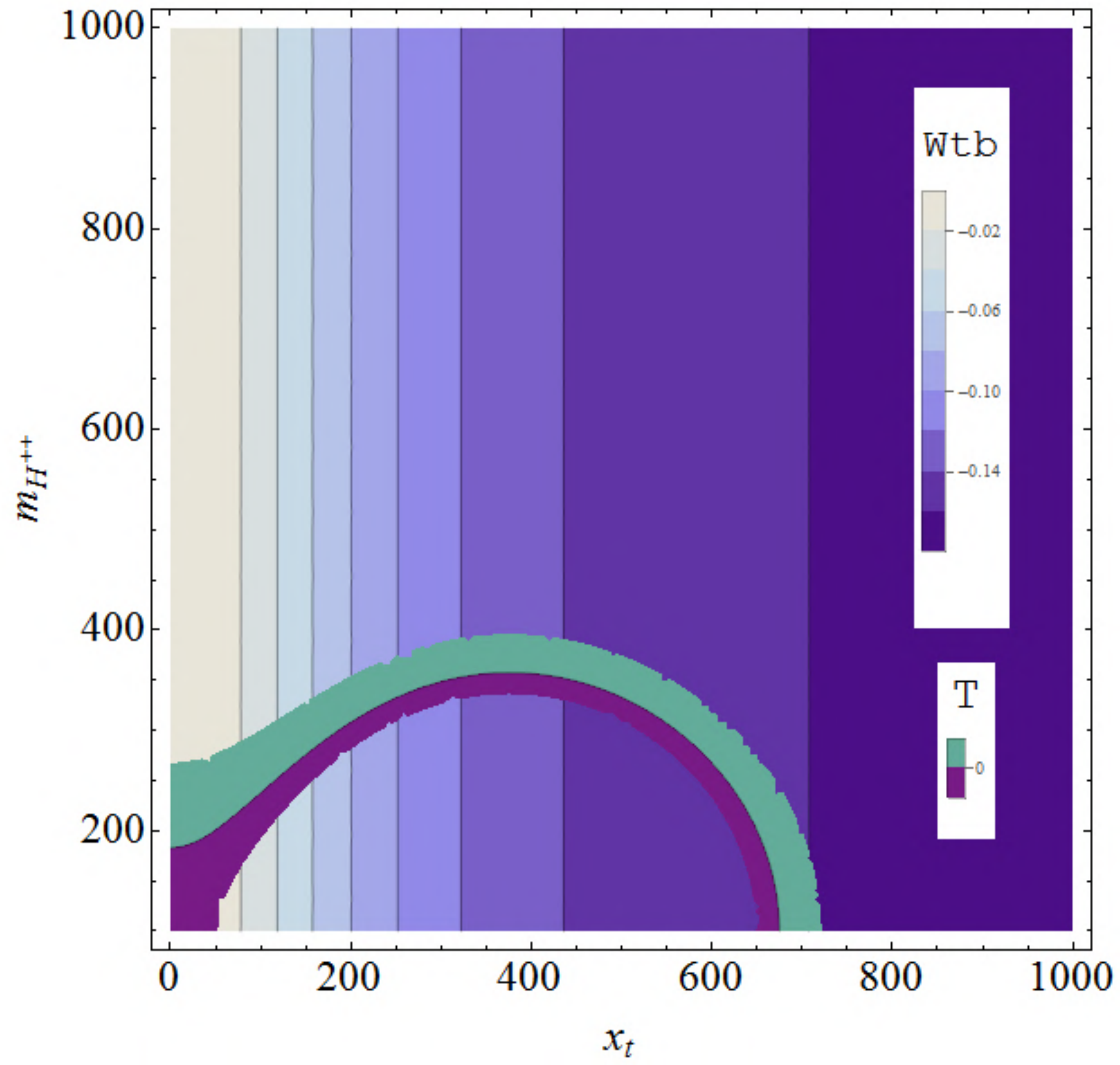}
&
    \includegraphics[width=3.8in,height=3.2in]{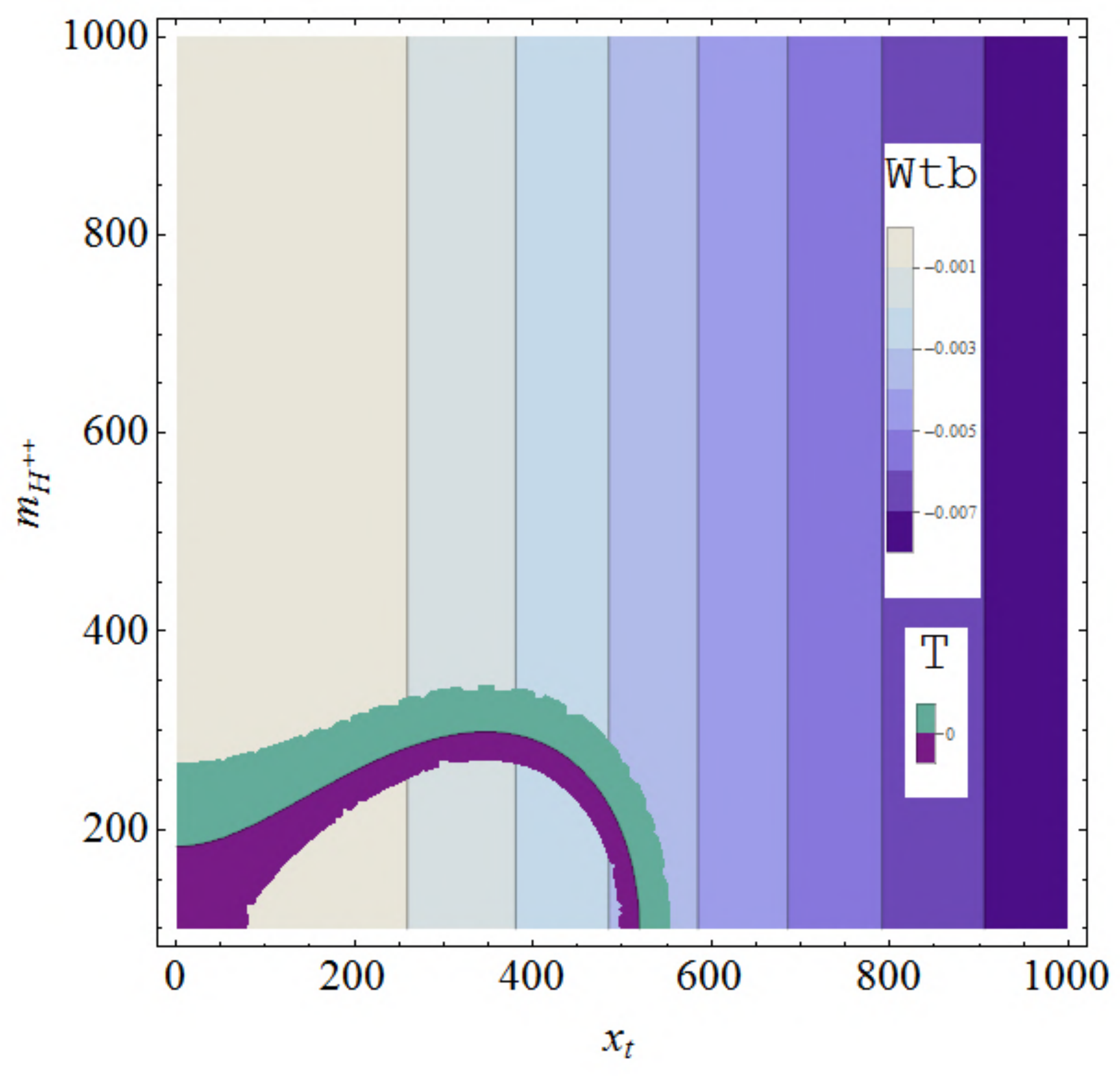}\\
    \hspace*{-1.7cm}
    \includegraphics[width=3.8in,height=3.2in]{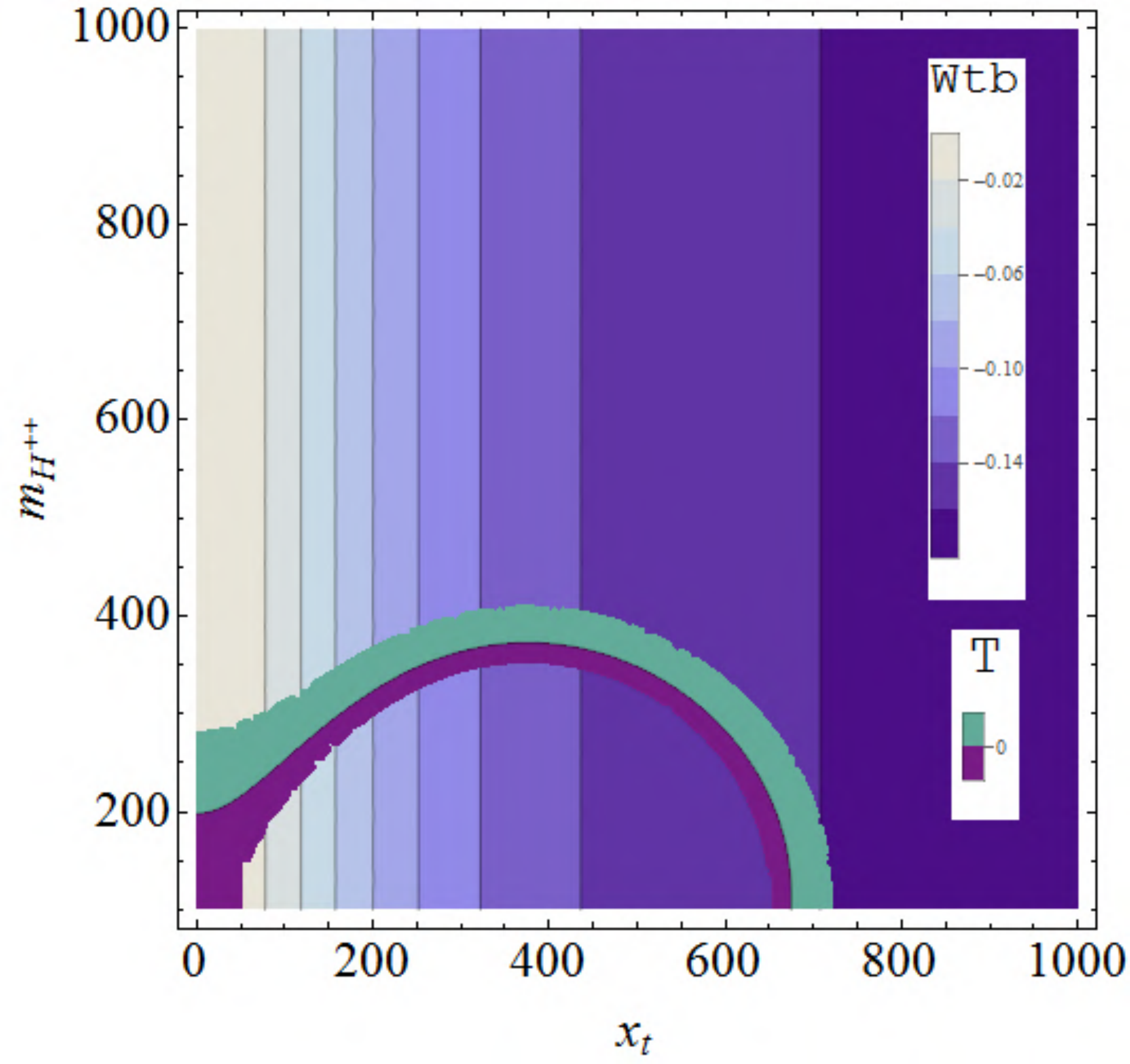}
&
    \includegraphics[width=3.8in,height=3.2in]{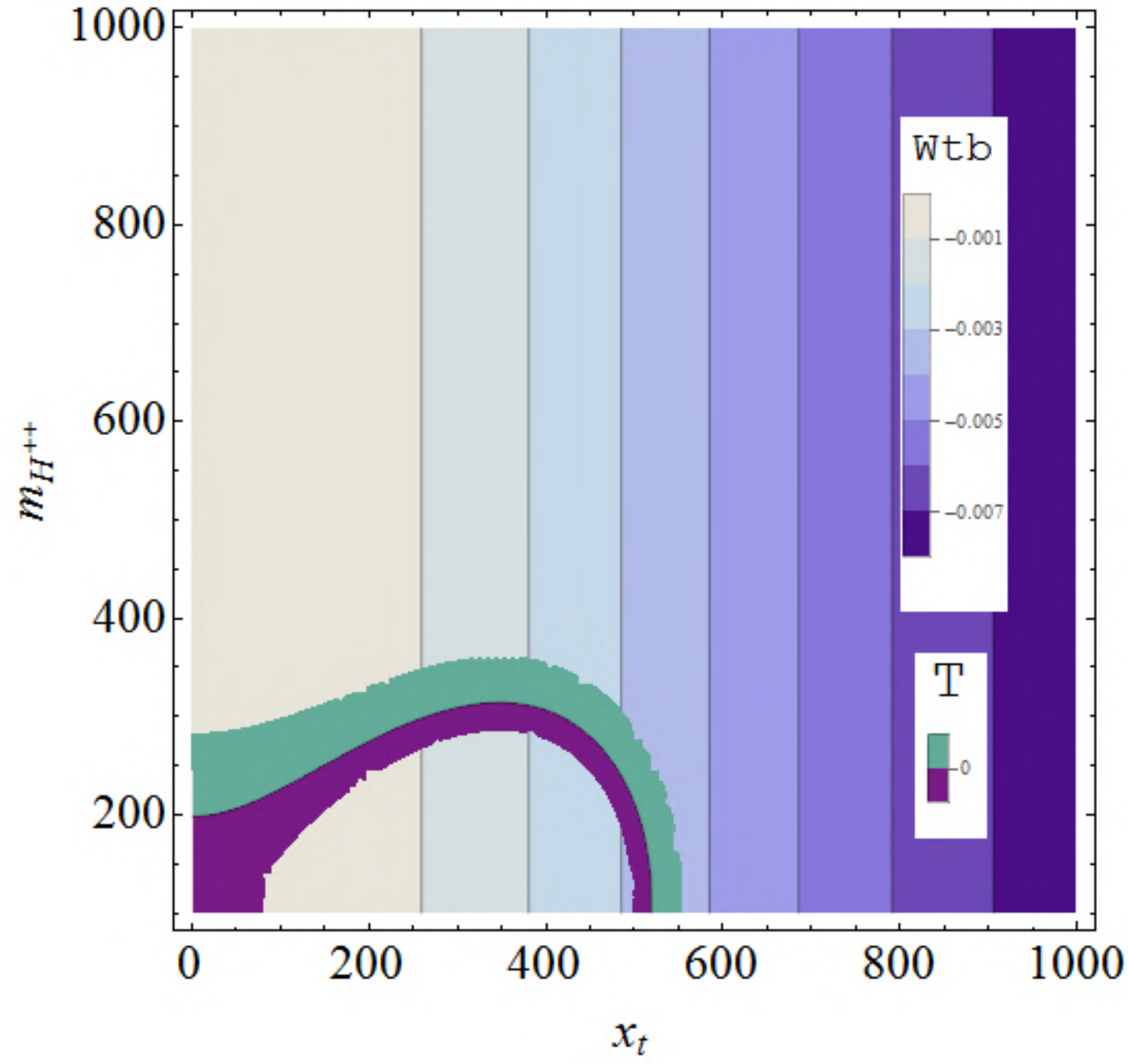}\\
        \end{array}$
        \end{center}
\vskip -0.1in
     \caption{(color online). \sl\small Same as Fig. \ref{fig:MTinD1HTM}, but for the ${\cal D}_X$ scenario.}
\label{fig:MTinDXHTM}
\end{figure}

Lower bounds on the masses of the vector-like quarks  have been obtained under various scenarios \cite{Carmi:2012yp,Ellis:2014dza,Aguilar-Saavedra:2013qpa}. But in the HTM, masses of these states are more restricted by electroweak constraints.   Fig. \ref{fig:MTinD1HTM} shows the dependence of $T$ parameter and its restriction as a contour plot in the  $m_{H^{\pm \pm}}-x_b$ plane for model ${\cal D}_1$, for light vector-like quark masses ($M=305$ GeV), in the left-hand side of the figure, and heavy vector-like masses ($M=1000$ GeV), in the right-hand side of the figure. The upper panels are for no mixing case, $\sin \alpha=0$, while the lower panels represent the maximal mixing case, $\sin \alpha=1$. In this case, the tree-level decay $Z \to b {\bar b}$ imposes a lower bound on the $x_b$ parameter, $x_b \ge 117$ GeV, for scenario ${\cal D}_1$, and this limit is the same from retractions on positive and negative deviations in $\delta R_{b}$. A similar plot for the ${\cal D}_X$ model in Fig. \ref{fig:MTinDXHTM} indicates that $Wtb$ does not impose similar restrictions on $x_t$.  

The mixing parameter $x_{b(t)}$ is also restricted by the $T$ parameter, as shown in Fig. \ref{fig:MTinD1HTM}, In the ${\cal D}_1$ model with vector-like quark mass $M=305$ GeV, the maximum $x_b$ allowed is $x_b=400$ GeV, while for $M=1000$ GeV, the maximum $x_b$ allowed is $x_b=538$ GeV. The same analysis for the ${\cal D}_X$ model (Fig. \ref{fig:MTinDXHTM}) shows that, with vector-like quark mass $M=305$ GeV, the upper limit for $x_t$ is $x_t=723$ GeV, while for $M=1000$ GeV, the upper limit for $x_t$ is $x_t=553$ GeV. 

We note that, Fig. \ref{fig:MTinD1HTM} and Fig. \ref{fig:MTinDXHTM} also indicate the constraints on the doubly charged mass, as a function of $x_{b(t)}$, for fixed values of vector-like quark mass parameter $M$,  from restrictions on the $T$ parameter.  In the ${\cal D}_1$ model, for $\sin \alpha=0$, the maximum doubly charged mass values allowed are $m_{H^{\pm\pm}}=382$ GeV and $m_{H^{\pm\pm}}=329$ GeV, reached for $M=305$ GeV and $M=1000$ GeV respectively, while for $\sin \alpha=1$, the maximum doubly charged mass values allowed are $m_{H^{\pm\pm}}=397$ GeV and $m_{H^{\pm\pm}}=343$, for $M=305$ GeV and $M=1000$ GeV respectively. In the plots for the ${\cal D}_1$ model we include restrictions from $Z b {\bar b}$ decay, which set lower limits on $x_b$ but seems not to affect $m_{H^{\pm\pm}}$.   
In the ${\cal D}_X$ scenario, for $\sin \alpha=0$, the maximum doubly charged mass values allowed are $m_{H^{\pm\pm}}=397$ GeV and $m_{H^{\pm\pm}}=345$ GeV, when $M=305$ GeV and $M=1000$ GeV, respectively, while for $\sin \alpha=1$, the maximum doubly charged mass values allowed are $m_{H^{\pm\pm}}=413$ GeV and $m_{H^{\pm\pm}}=360$ GeV, for $M=305$ GeV and $M=1000$ GeV, respectively.  Again the $Wtb$ vertex does not limit $x_t$ or $m_{H^{\pm\pm}}$.

To summarize, experimental constraints on $Wtb$ impose restrictions on $M$ in ${\cal D}_X$ scenario, while leaving $x_t$ free; while in the ${\cal D}_1$ scenario $Z b {\bar b}$ imposes restrictions on $x_b$ while leaving $M$ unconstrained.

\section{Effect of vector quarks on $H \to \gamma \gamma$ and $H \to Z \gamma$}
\label{sec:hgg}
The production and decays of the vector-like quarks will proceed in the same way as in the SM, and this was explored extensively before. However, what could be different are effects on the loop-induced decays $H \to \gamma \gamma$ and $H \to Z \gamma$, through interplays of contributions of additional particles in the loop, in our case charged and doubly charged Higgs bosons and vector-like quarks. So in this section, we study vector-like quarks contribution to the Higgs decay in the HTM.  The decay width $h \to \gamma \gamma$ is
\begin{eqnarray}
\label{eq:THM-h2gaga}
[\Gamma(h \rightarrow\gamma\gamma)]_{HTM}
& = & \frac{G_F\alpha^2 m_{h}^3}
{128\sqrt{2}\pi^3} \bigg| \sum_{f} N^f_c Q_f^2 g_{h ff} 
A_{1/2}
(\tau^h_f) + g_{h WW} A_1 (\tau^h_W) + \tilde{g}_{h H^\pm\,H^\mp}
A_0(\tau^h_{H^{\pm}}) \nonumber \\
&+& 
 4 \tilde{g}_{h H^{\pm\pm}H^{\mp\mp}}
A_0(\tau^h_{H^{\pm\pm}})   + \sum_{q}\frac{Y_{qq} Q_q^2 N^f_c g_{hff}}{m_{q}} A_{1/2}(\tau^h_{q})       \bigg|^2 \, ,
\label{partial_width_htm}
\end{eqnarray}
where the sum runs over $q= t,~T$ for up-type quarks and over $b,~B$ for down-type ones. The value for $m_{T}$ is given in Eq. (\ref{eq:eigenvalues}), and the loop functions for spin $0$, spin $1/2$ and spin $1$ appear in the literature. For this, and for the couplings of $h$ to the vector bosons and fermions, and the scalar trilinear couplings we use the same expressions as in our previous work \cite{Bahrami:2013bsa}. 
The couplings of the Higgs bosons with vector quarks ($Y_{qq}$) appearing in Eq. (\ref{partial_width_htm}) are listed in Appendix.

The new quarks effect on the di-photon search channel at the LHC is expressed  by the ratio
\begin{eqnarray}
\label{eq:Rgamgam}
R_{\gamma \gamma}=\frac{[\sigma( gg \to  h) \times \Gamma(h \to \gamma \gamma)]_{HTM}}{[\sigma( gg \to\Phi) \times \Gamma(\Phi \to \gamma \gamma)]_{SM}} \times \frac{[\Gamma( \Phi)]_{SM}}{[\Gamma(h)]_{HTM}}, 
\end{eqnarray}
where $\Phi$ is the SM neutral Higgs boson. We neglect the contribution of the $b$ quark.  The ratio of the production cross sections by gluon fusion is
\begin{equation}
\frac {\sigma_{\rm HTM}(gg \to h )}{\sigma_{\rm SM}(gg \to  \Phi )}=\left[  g_{hff}+\frac{\sum_{q} \displaystyle \frac{Y_{qq} g_{hff}}{m_{q}} A_{1/2}(\tau^h_{q})}
{A_{1/2}(\tau^h_{t})} \right]^2 \, .
\end{equation}
 
 As in \cite{Arbabifar:2012bd},  we set the values 125 GeV and 98 GeV for the $h$ and $H$ masses respectively, and adjust the parameters $\lambda_1-\lambda_5$ accordingly.
The relative widths factor is as defined  in \cite{Bahrami:2013bsa}.
Previously,  in \cite{Wang:2012gm}, couplings of the vector quarks and in the Higgs potential were assumed to be arbitrary, and thus the $gg \to H$ production rate could be reduced to 20\% of the SM value.  In our considerations,  vector quark couplings are restricted from the mixing matrices Eq. (\ref{eq:mixing_s,t}) and Eq. (\ref{eq:mixing_d}), and we relate the couplings in the Higgs potential to the Higgs masses \cite{Arbabifar:2012bd}. 

Our numerical investigations agree with those in \cite{Aguilar-Saavedra:2013qpa}. In both the loops for Higgs production through gluon fusion, and in the loops for Higgs di-photon decay, the contributions of the vector -like quarks are very small. This effect is stronger than expected by decoupling, and arise also from small couplings of the new quarks, given in the Appendix. The couplings of the new quarks to the Higgs bosons is limited by the trace of the mixing matrix for both singlet/triple and for doublet representations, which must equal 1 \cite{Aguilar-Saavedra:2013qpa}. Even for light masses, varying $\displaystyle M \in (100-500)$~GeV and $\displaystyle x_{b(t)} \in (0-1000)$~GeV, the variation in $R_{\gamma \gamma}$ is less than 10\%, and thus below the precision of the current measurements at the LHC.

The decay rates $R_{\gamma \gamma}$ and $R_{Z \gamma}$ depend sensitively on $\sin \alpha$ and $m_{H^{\pm\pm}}$. We investigate this dependence in the context of the HTM model with vector quarks, because, although the vector quarks do not explicitly modify the di-photon and Z-photon decays, they affect the parameter space of $\sin \alpha$ - $m_{H^{\pm\pm}}$ through restrictions on the $T$ parameter, and thus they {\it indirectly} affect the decays.

\begin{figure}[htbp]
\begin{center}$
    \begin{array}{ccc}
\hspace*{-1.7cm}
    \includegraphics[width=2.5in,height=2.8in]{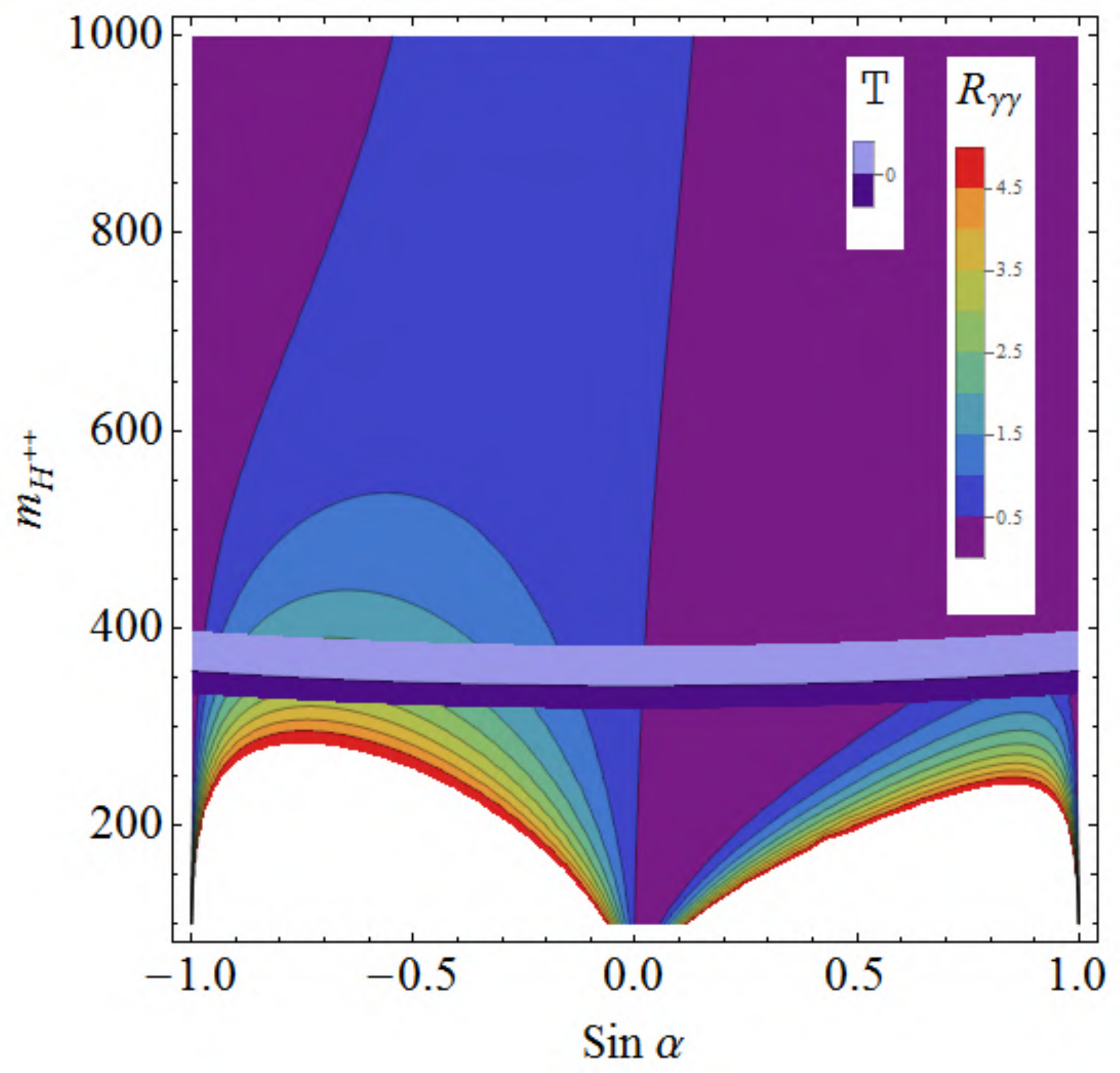}
&\hspace{-0.2cm}
\includegraphics[width=2.5in,height=2.8in]{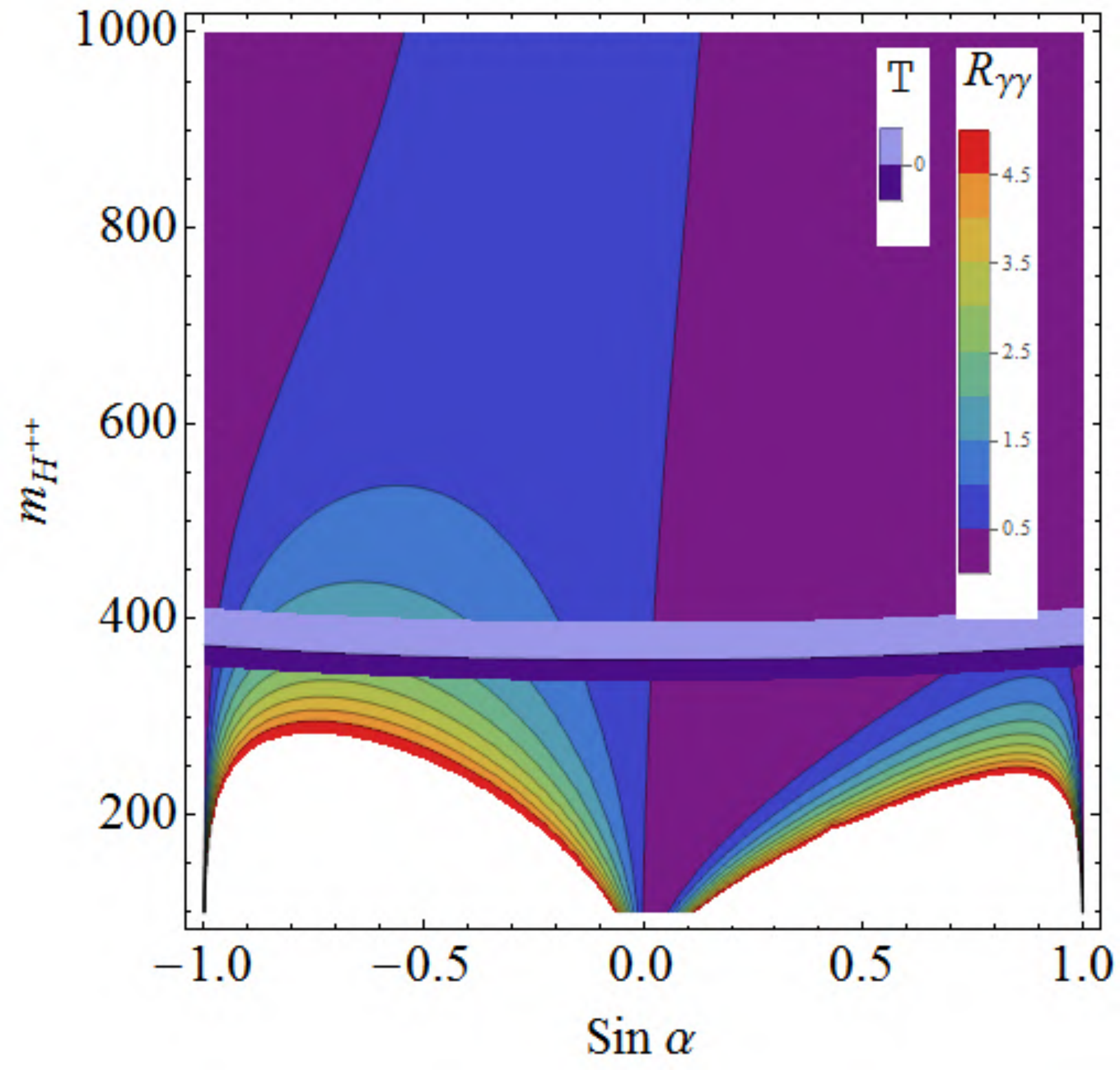}                                                                             
& \hspace{-0.2cm}
\includegraphics[width=2.5in,height=2.8in]{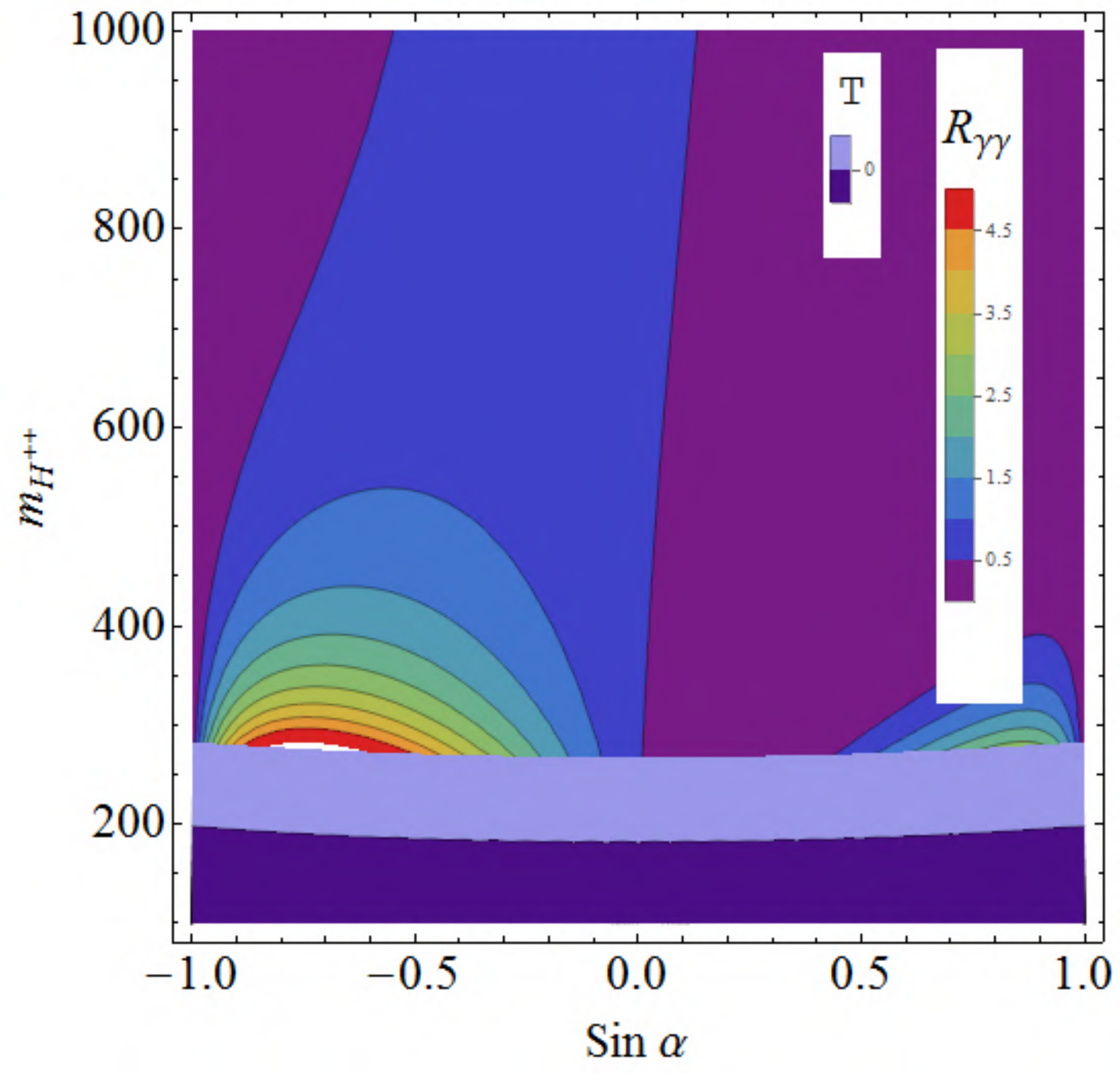}\\
    \hspace*{-1.7cm}
    \includegraphics[width=2.5in,height=2.8in]{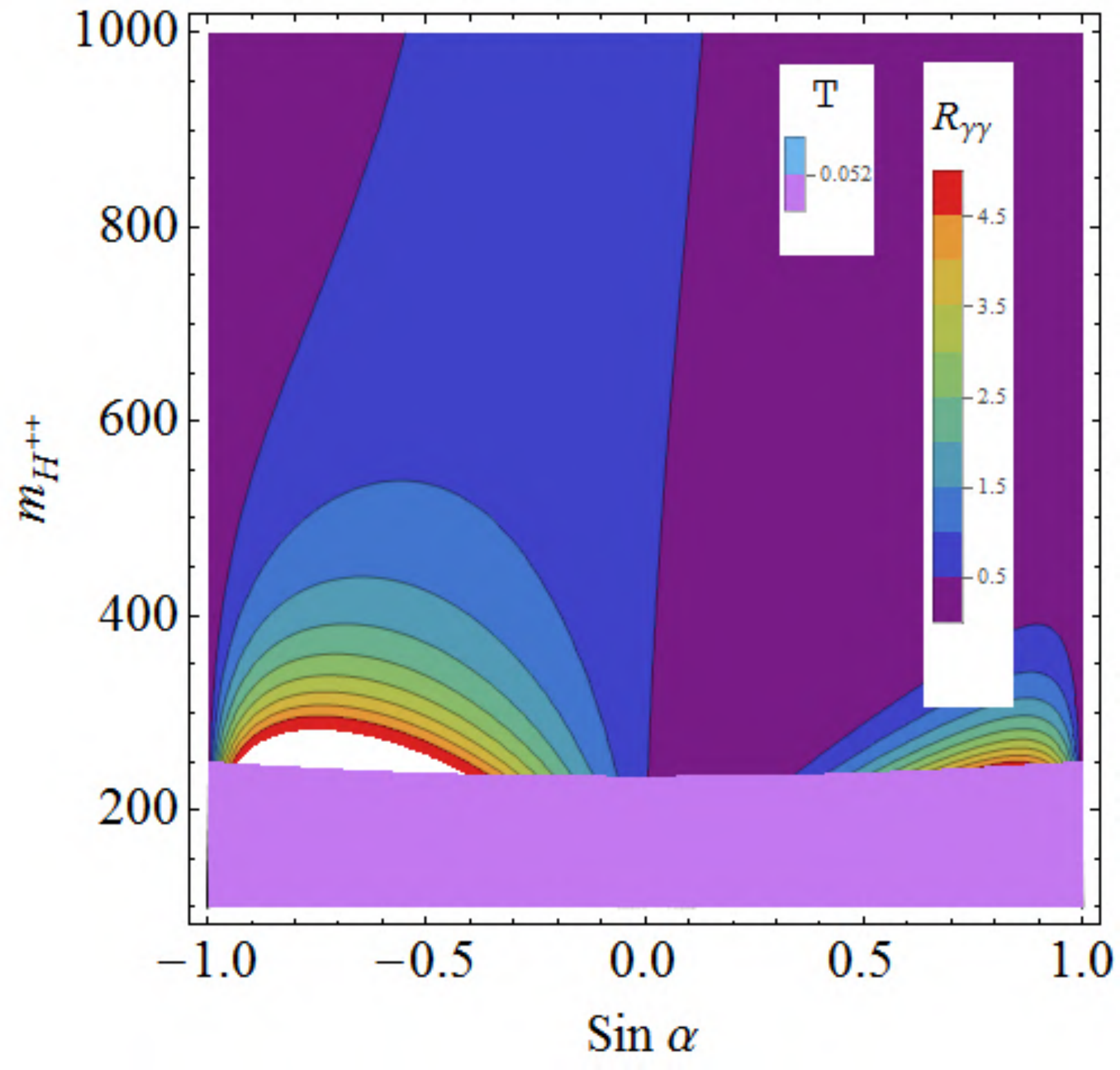}
&\hspace{-0.2cm}
\includegraphics[width=2.5in,height=2.8in]{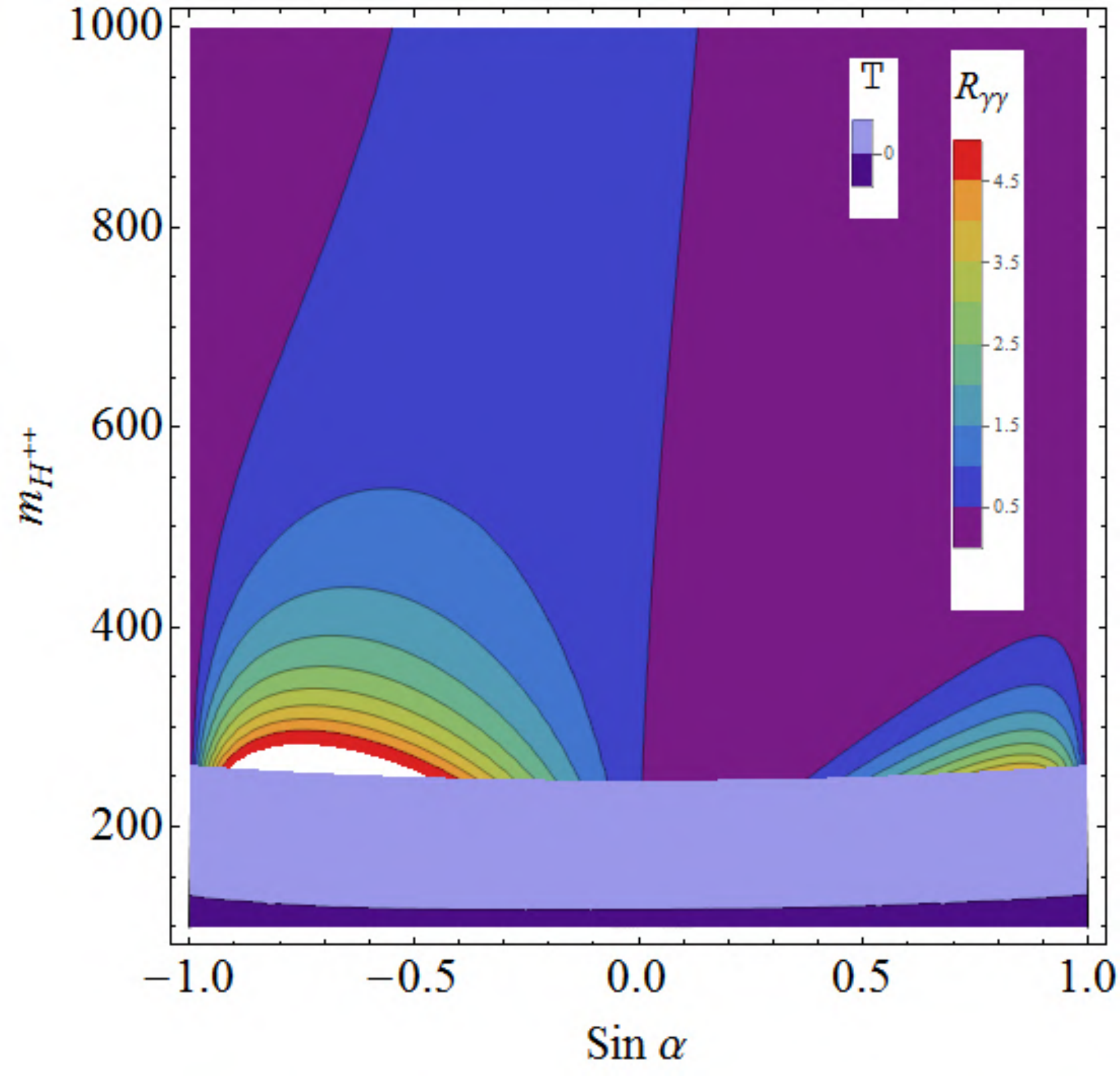}
   & \hspace{-0.2cm}
   \includegraphics[width=2.5in,height=2.8in]{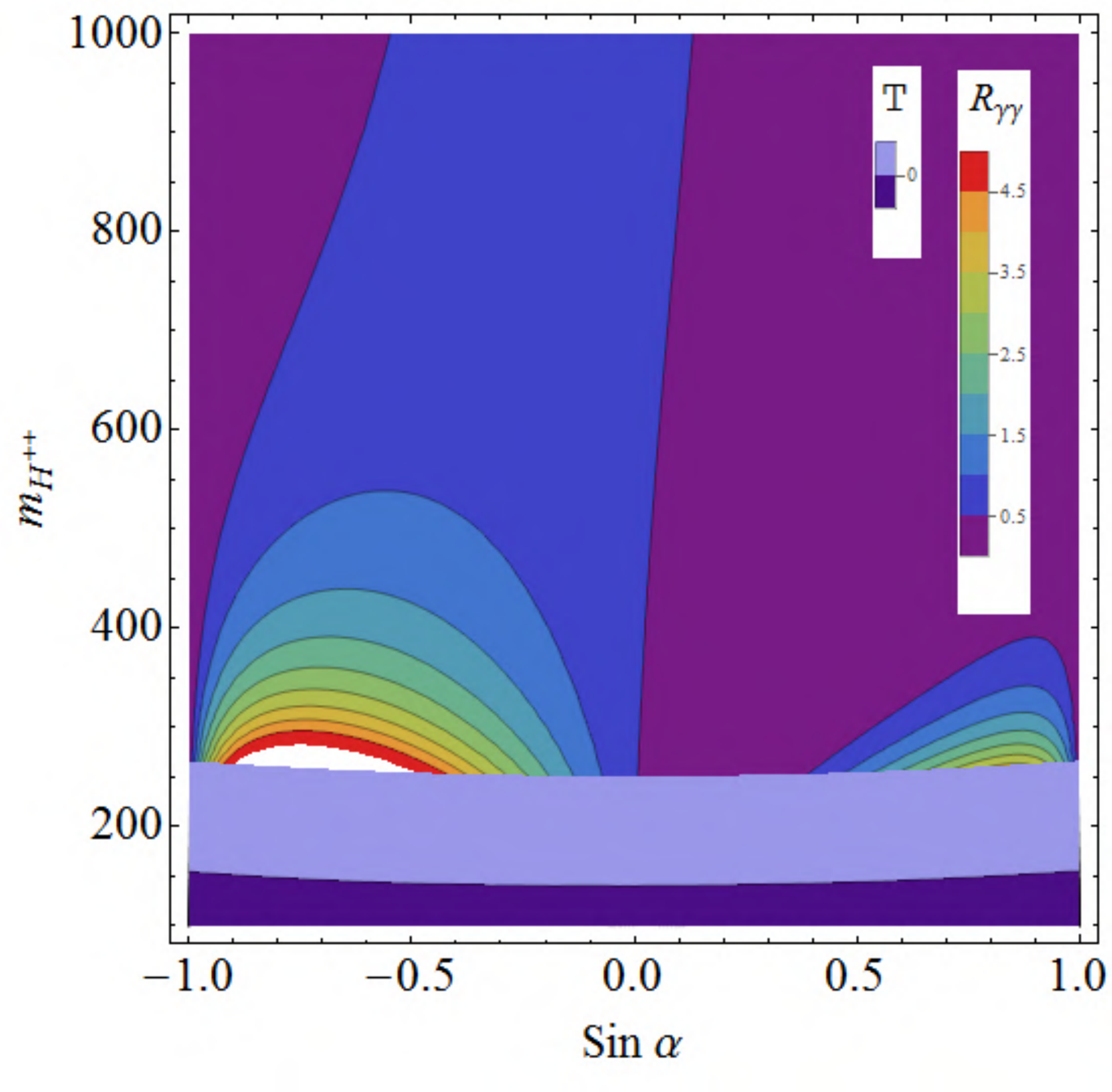}\\
        \end{array}$
\end{center}
\vskip -0.1in
     \caption{(color online). \sl\small Contour graphs for the relative strength of the Higgs di-photon decay $R_{\gamma \gamma}$, including the restrictions of the $T$  parameter in the HTM with vector quarks, as a function of the doubly charged Higgs boson mass $m_{H^{\pm \pm}}$ and the mixing in the CP-even neutral Higgs sector, $\sin \alpha$. We show plots for scenario ${\cal D}_1$ (upper left panel), scenario ${\cal D}_X$ (upper middle panel),  scenario ${\cal D}_Y$ (upper right panel), scenario ${\cal T}_X$ (lower left panel), scenario ${\cal D}_2$ (middle lower panel) and scenario ${\cal U}_1$ (lower right panel).  Results for scenario ${\cal T}_Y$ are not shown, but summarized in the text. We took $M=305$ GeV for all the graphs, and values for $x_t, x_b$ consistent with a larger allowed range for $m_{H^{\pm \pm}}$.}
\label{fig:Rgg}
\end{figure}

The results of our analyses are shown in Fig. \ref{fig:Rgg}. In purple, we draw contour plots for the $T$ parameter restrictions, while values for $R_{\gamma \gamma}$ are shown in multicolor contours. We have drawn plots for scenarios (in order, from the top,  left to right side): ${\cal D}_1$, ${\cal D}_X$, ${\cal D}_Y$, ${\cal T}_X$, ${\cal D}_2$,  ${\cal U}_1$, but we omit plots for  for scenario ${\cal T}_Y$, for brevity, and because for this model the allowed range for $m_{H^{\pm \pm}}$ for the parameters chosen is the smallest. The differences in the contours for $R_{\gamma\gamma}$ between models are negligible: however, what differs amongst models are restrictions on the values of the doubly charged Higgs boson mass. 

For the model ${\cal D}_1$  (top left panel), $R_{\gamma \gamma}$ can take values between $0.5$ and  $4$, but the mass $m_{H^{\pm \pm}}$ is restricted to lie in a band $(318-400)$ GeV; while  for scenario ${\cal D}_X$  (middle top panel), $R_{\gamma \gamma}$ can take values between $0.5$ and  $3$, but the mass $m_{H^{\pm \pm}}$ is restricted to lie in a band $(333-412)$ GeV.  For the other scenarios, $R_{\gamma \gamma}$ can take values between $0.5$ and  $5$, but the mass $m_{H^{\pm \pm}}$ is restricted to lie in a band $(100-263)$ GeV for the ${\cal D}_2$ model, in $(100-283)$ GeV for the ${\cal D}_Y$ model, in $(100-250)$ GeV for the ${\cal T}_X$ model, and in $(100-268)$ GeV for the ${\cal U}_1$ model. The restriction on the mass of the doubly charged Higgs boson is thus what differentiates these models.  

We note also that, as in HTM without additional fermions, for $\sin \alpha=0$ the Higgs di-photon decay {\it cannot} be enhanced with respect to its SM value. This confirms our previous analyses in \cite{Arbabifar:2012bd, Bahrami:2013bsa}.   The relative branching ratios $R_{\gamma \gamma}$ are very sensitive to values of $\sin \alpha$. In the allowed regions of $m_{H^{\pm \pm}}$ bands, the angle for which the enhancement in the di-photon decay is $1.5 - 3.5$ times the SM value is $\sin \alpha \in (-0.95, -0.23)$ in model ${\cal D}_1$, while in model ${\cal D}_X$ the same enhancement is obtained for $\sin \alpha \in (-0.93, -0.25)$. In this model, an enhancement is also possible for positive $\sin \alpha \in (0.75, 0.93)$, where $R_{\gamma \gamma} =1.5$.  In the ${\cal D}_2$ model an enhancement of $R_{\gamma \gamma}$ of $1.5-3.5$ is obtained for a large range of both positive $\sin \alpha \in (0.08, 1)$ and negative values $\sin \alpha \in (-1, -0.02)$, and the same holds for the other models, ${\cal D}_Y$, ${\cal T}_X$,  ${\cal U}_1$ and ${\cal T}_Y$. As a general feature, $R_{\gamma \gamma}$ is more enhanced at negative values of $\sin \alpha$. In all the plots we chose a light $M=305$ GeV, just above our required minimum, and values for $x_t$ and $x_b$ consistent with a large allowed parameter range for $m_{H^{\pm \pm}}$. For models ${\cal D}_1$, the plots are for $x_b=230$ GeV, above the required minimum, while for model ${\cal D}_X$, the plots are for $x_t=370$ GeV, consistent with the previous section. For other models, the restrictions for $x_b$ and $x_t$ are much relaxed, and we have chosen $x_b=20$ GeV and $x_t=50 GeV$, as in previous studies \cite{Cai:2012ji,Cacciapaglia:2010vn}. An exception is model ${\cal T}_Y$, where in order to have a $-0.2 \le \Delta T \le 0.4$, $x_t < 30$ GeV, and the mass range for the doubly charged Higgs boson increases with decreasing $x_t$.

In general, for heavier $M$ and lighter $x_{t(b)}$, we can obtain a slightly higher upper band for mass of doubly charged Higgs bosons. The exceptions are, the ${\cal T}_Y$ model, as mentioned above, and ${\cal  D}_1$ and ${\cal D}_X$ models,  where higher upper limits for $m_{H{pm \pm}}$ are obtained for lighter vector-like quark masses $M$.

The decay width for $h \to Z \gamma$ is given by \cite{Chen:2013vi}:
\begin{eqnarray}
\label{eq:hzg}
[\Gamma(h \rightarrow Z\gamma)]_{HTM}
& = & \frac{\alpha G_F^2  m_W^2 m_{h}^3}
{64\pi^4} \left ( 1-\frac{m_Z^2}{m_h^2} \right )^3  \bigg|\frac{1}{c_W}  \sum_{f} 2 N^f_c Q_f (I_3^f-2Q_f s_W^2) g_{h ff} 
A^h_{1/2}
(\tau^h_f, \tau^Z_f)  \nonumber \\
 &+&\frac{1}{c_W}\sum_{q} 2 N^f_c Q_q (I_3^q-2Q_q s_W^2) \frac{Y_{qq}}{m_q} g_{h ff} 
A^h_{1/2}
(\tau^h_q, \tau^Z_q) \nonumber \\
&+& c_W g_{h WW} A^h_1 (\tau^h_W, \tau^Z_W)- 2s_W \tilde{g}_{h H^\pm\,H^\mp}g_{ZH^\pm H^\mp}
A^h_0(\tau^h_{H^{\pm}}, \tau^Z_{H^\pm}) \nonumber \\
&-&
 4s_W \tilde{g}_{h H^{\pm\pm}H^{\mp\mp}} {g}_{Z H^{\pm\pm}H^{\mp\mp}}
A^h_0(\tau^h_{H^{\pm\pm}}, \tau^Z_{H^{\pm \pm}})        \bigg|^2 \, ,
\end{eqnarray}
where the sum runs over $q= t,~T$ for up-type quarks and over $b,~B$ for down-type ones and  $\tau^h_i=4m_i^2/m_h^2,$ $\tau^Z_i=4m_i^2/m_Z^2$, with $i=t, T, b, B, W,H^\pm,H^{\pm\pm}$.  $I_3^f=\pm \frac12$ is the weak isospin of top and bottom quarks, while for vector-like quarks $I_3^{F}=I_3^f +f_L=f_R$, with $F=T,B$, and $f_L,~f_R$ depend on the vector-like quark representation \cite{Okada:2012gy}, and are listed in Table \ref{tab:fLfR}.
\begin{table}[htbp]
\caption{\label{tab:fLfR}\sl\small Neutral current parameters $f_L$ and $f_R$ for vector-like quarks $Z$ interaction.}
  \begin{center}
 \small
 \begin{tabular*}{0.99\textwidth}{@{\extracolsep{\fill}} c| cccccccc}
 \hline\hline
	Name &&${\cal U}_1$ &${\cal D}_1$ &${\cal D}_2$ &${\cal D}_X$ &${\cal D}_Y$ &${\cal T}_X$ 
	&${\cal T}_Y$\\
  Type&&Singlet &Singlet &Doublet&Doublet &Doublet &Triplet 
	&Triplet \\
	 \hline
	   $T$ & $\begin{array}{c} f_L \\ f_R \end{array} $& 
$\begin{array}{c} -1/2 \\  0   \end{array}$ &  & 
$\begin{array}{c}    0 \\ +1/2 \end{array} $& 
$\begin{array}{c}  -1   \\ -1/2 \end{array} $&  &
$\begin{array}{c} -1/2 \\  0   \end{array} $ & 
$\begin{array}{c} +1/2 \\ +1   \end{array} $ \\
\hline
$B$ & $\begin{array}{c} f_L \\ f_R \end{array} $&  &
$ \begin{array}{c} +1/2 \\ 0    \end{array} $& 
$ \begin{array}{c}  0   \\ -1/2 \end{array} $& &
$ \begin{array}{c} +1   \\  1/2 \end{array} $& 
$ \begin{array}{c} -1/2 \\ -1   \end{array} $& 
$ \begin{array}{c} +1/2 \\  0   \end{array} $\\
\hline 
\hline
   \end{tabular*}
\end{center}
 \end{table}

The loop-factors are and couplings have been given before, and we use the expressions in \cite{Bahrami:2013bsa}.

\begin{figure}[htbp]
\begin{center}$
    \begin{array}{ccc}
\hspace*{-1.7cm}
    \includegraphics[width=2.5in,height=2.8in]{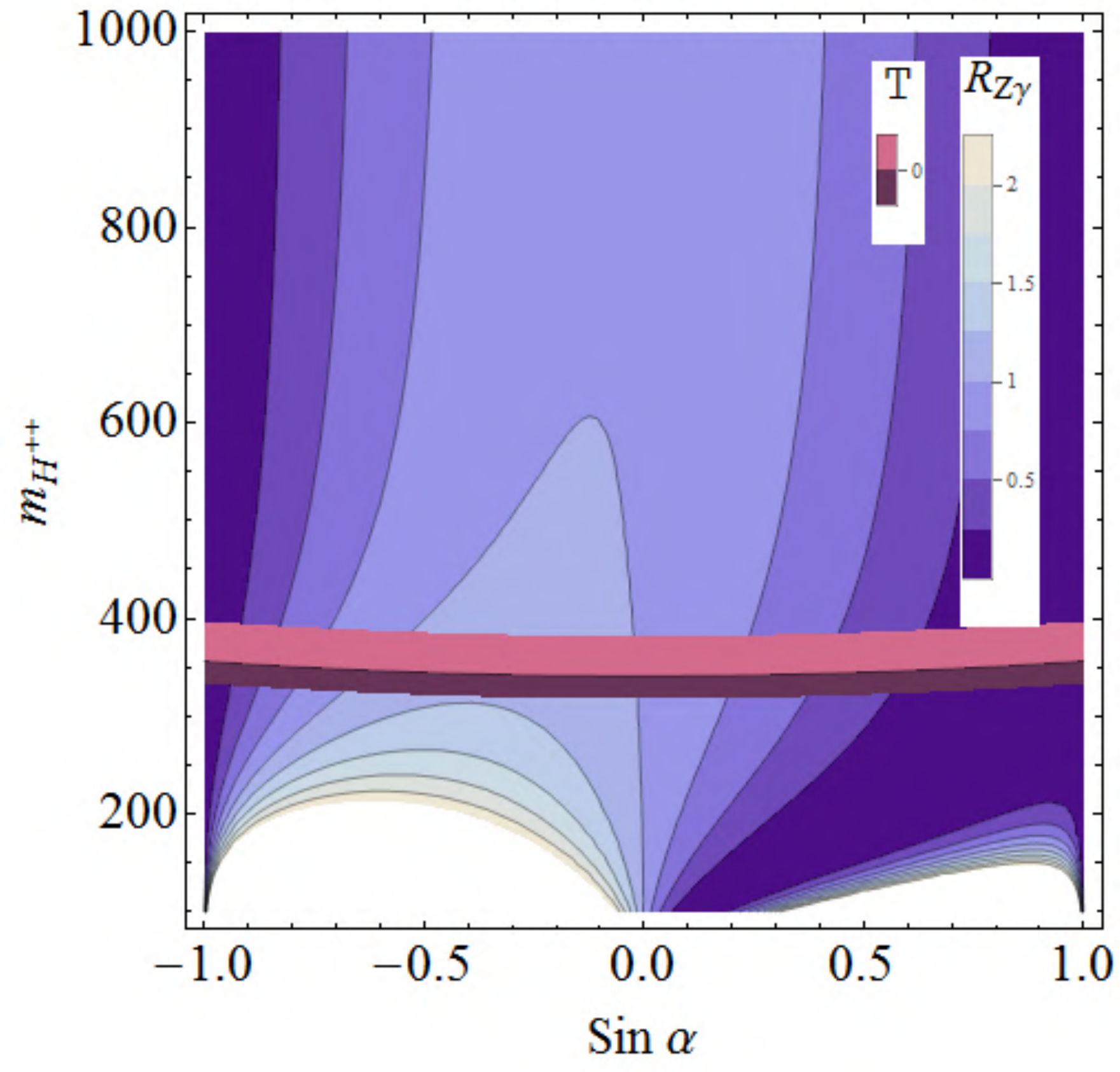}
&\hspace{-0.2cm}
\includegraphics[width=2.5in,height=2.8in]{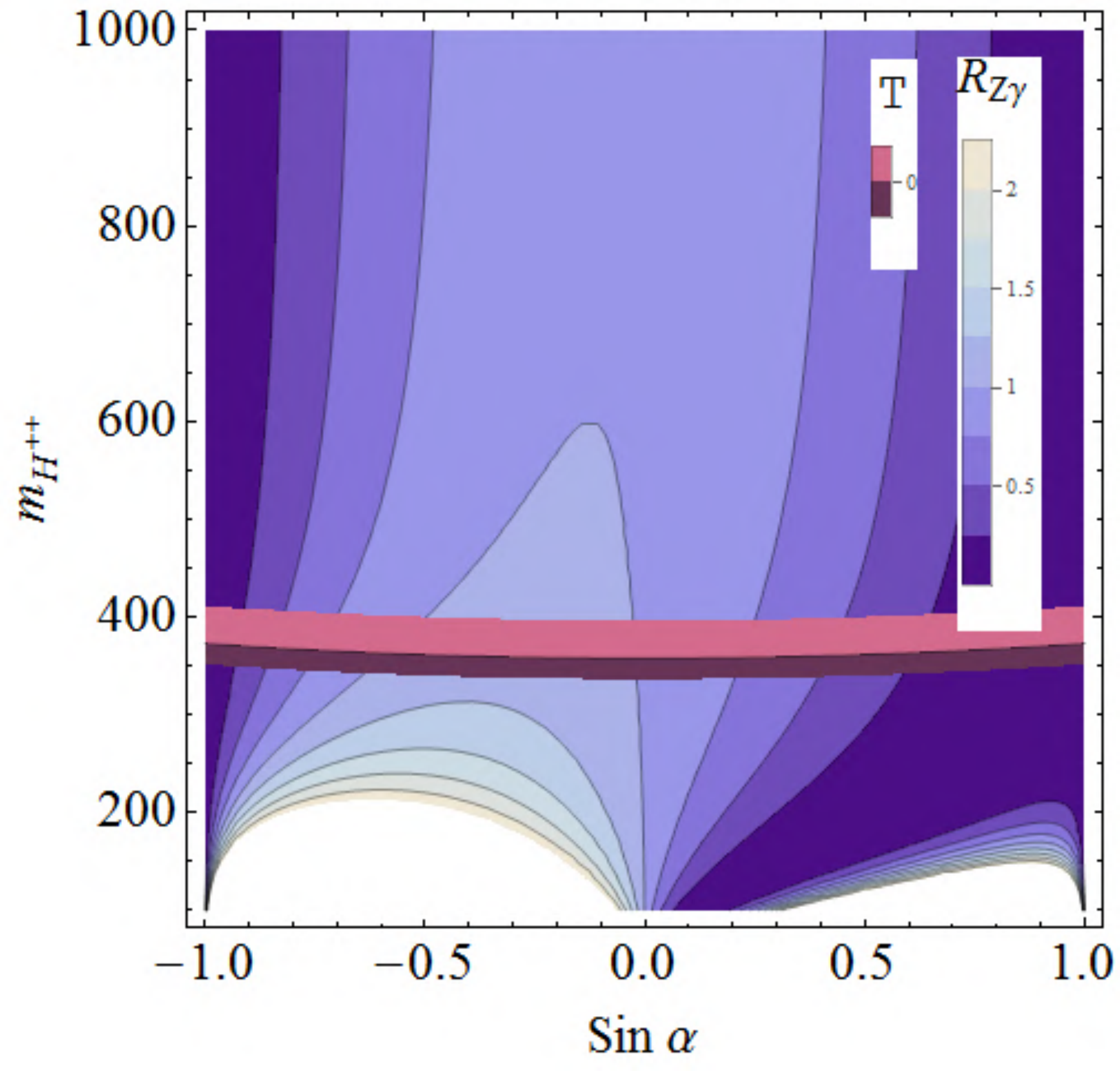}                                                                             
& \hspace{-0.2cm}
\includegraphics[width=2.5in,height=2.8in]{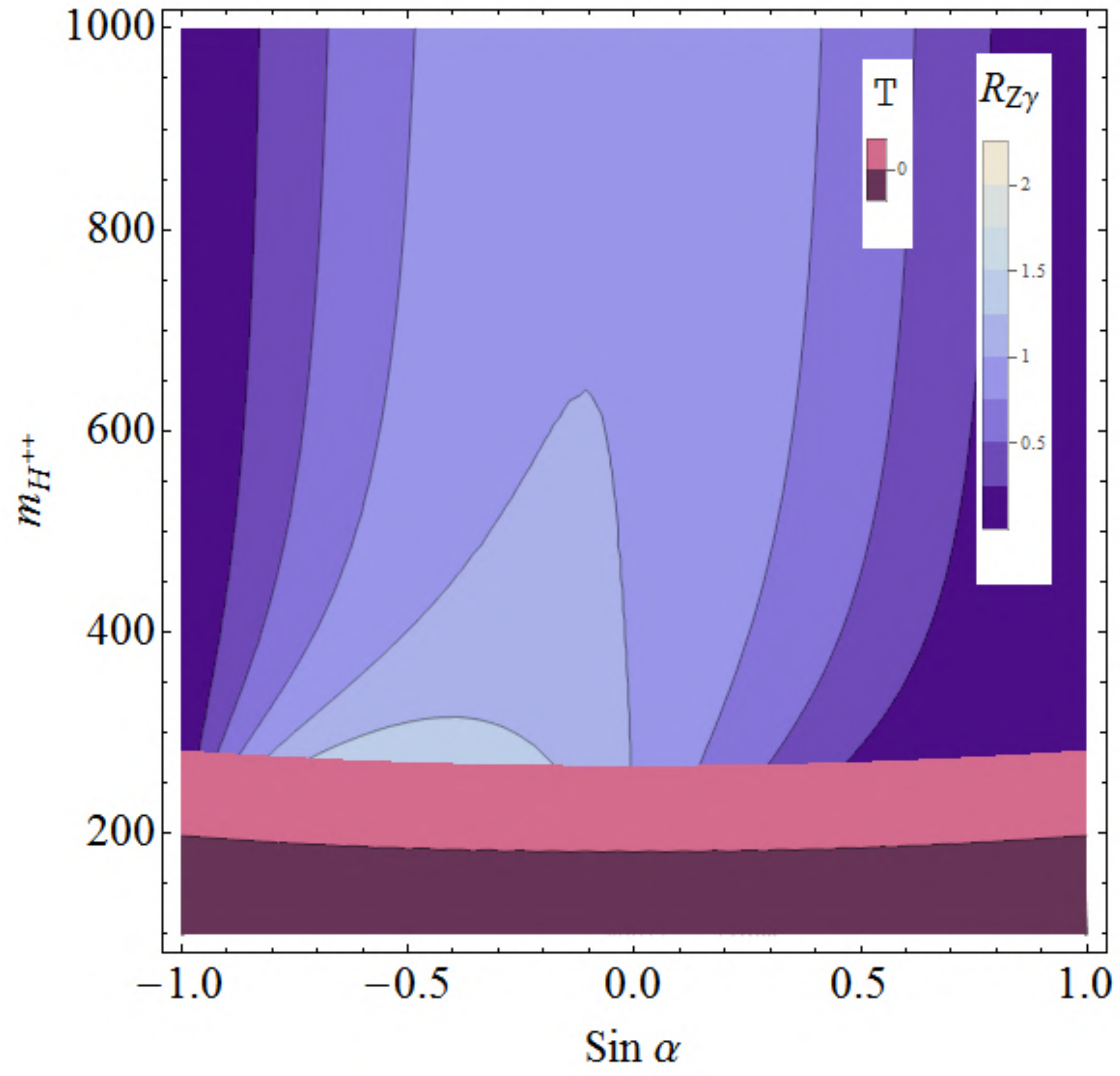}\\
    \hspace*{-1.7cm}
    \includegraphics[width=2.5in,height=2.8in]{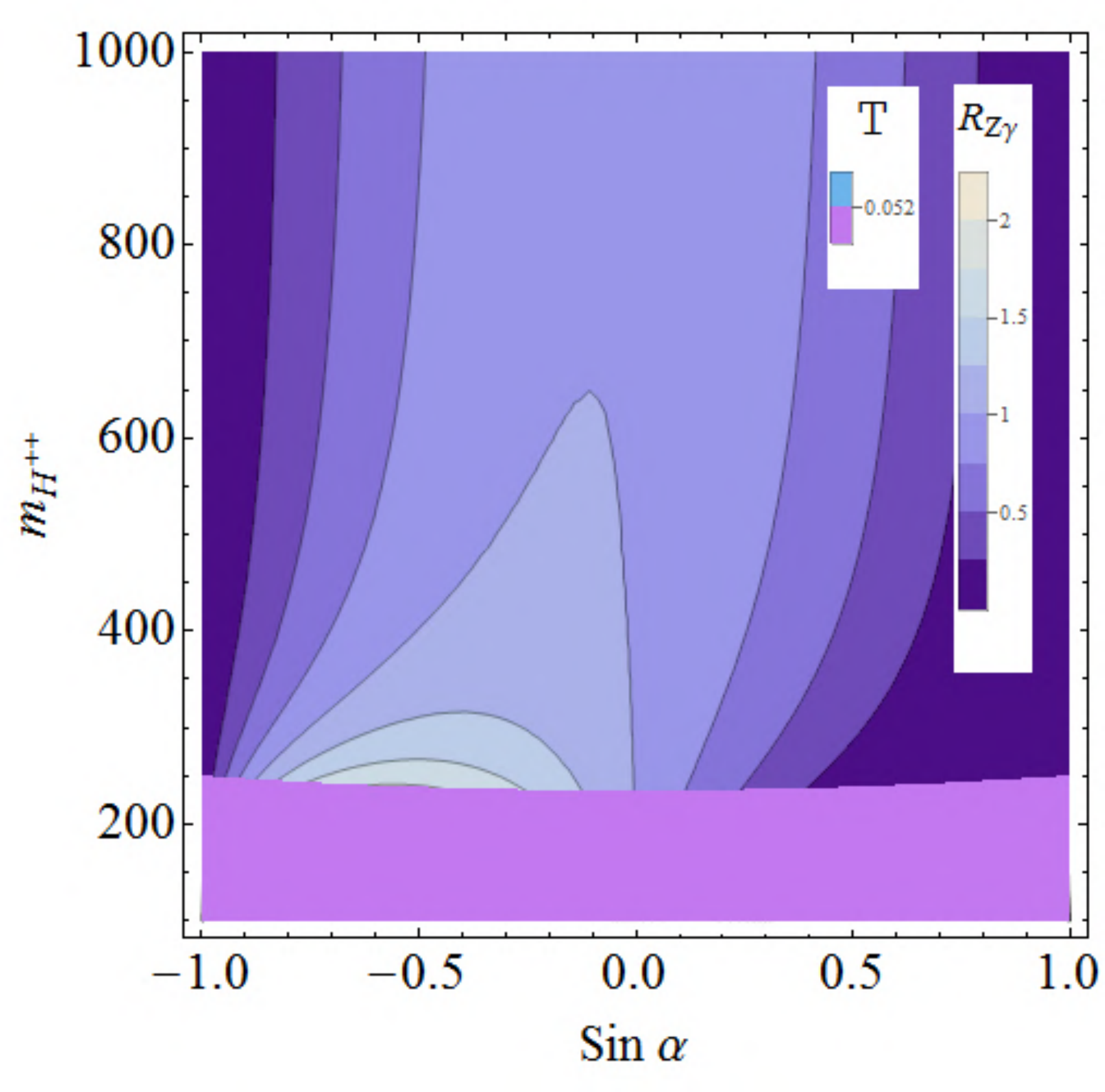}
&\hspace{-0.2cm}
\includegraphics[width=2.5in,height=2.8in]{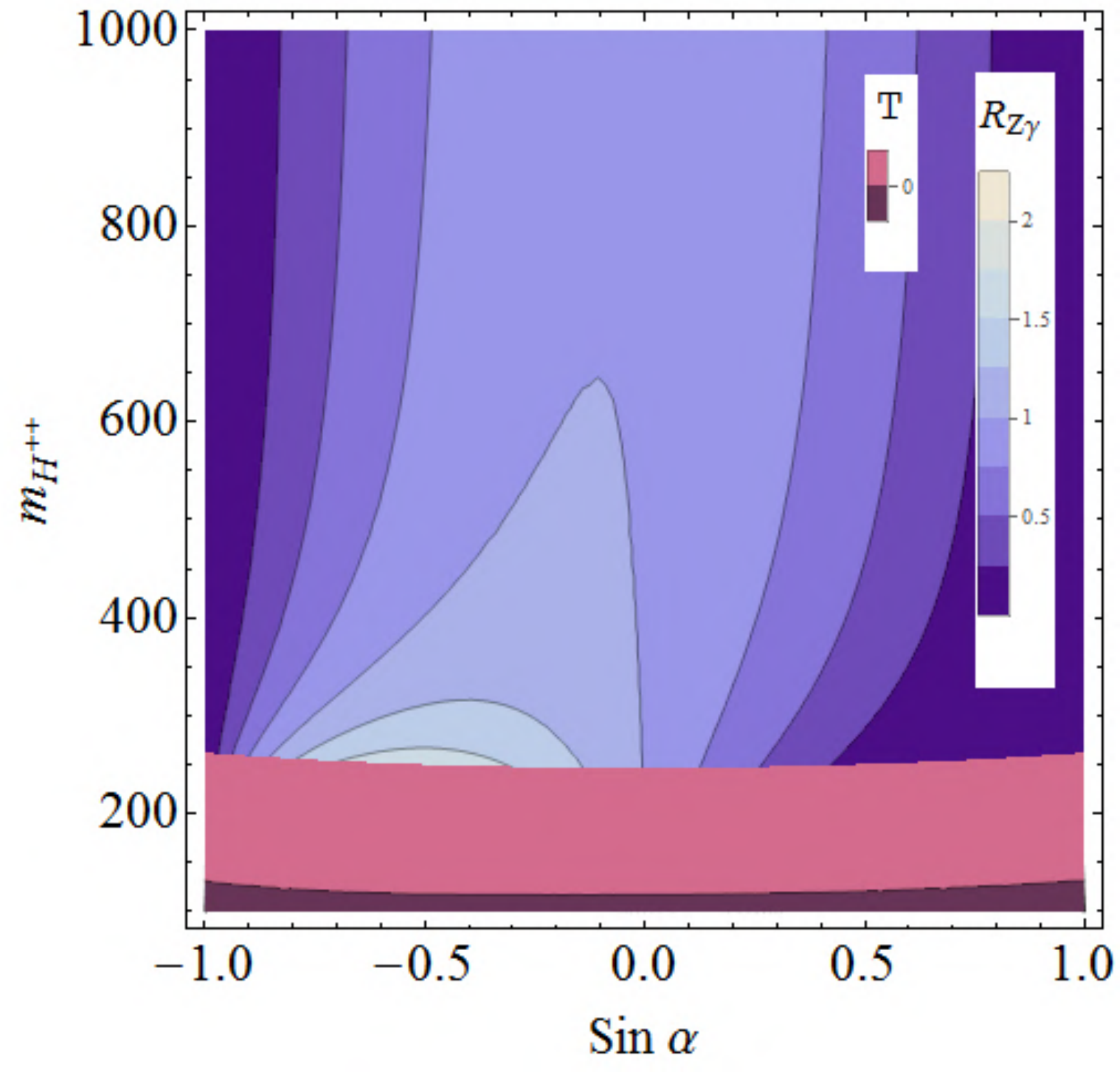}
   & \hspace{-0.2cm}
   \includegraphics[width=2.5in,height=2.8in]{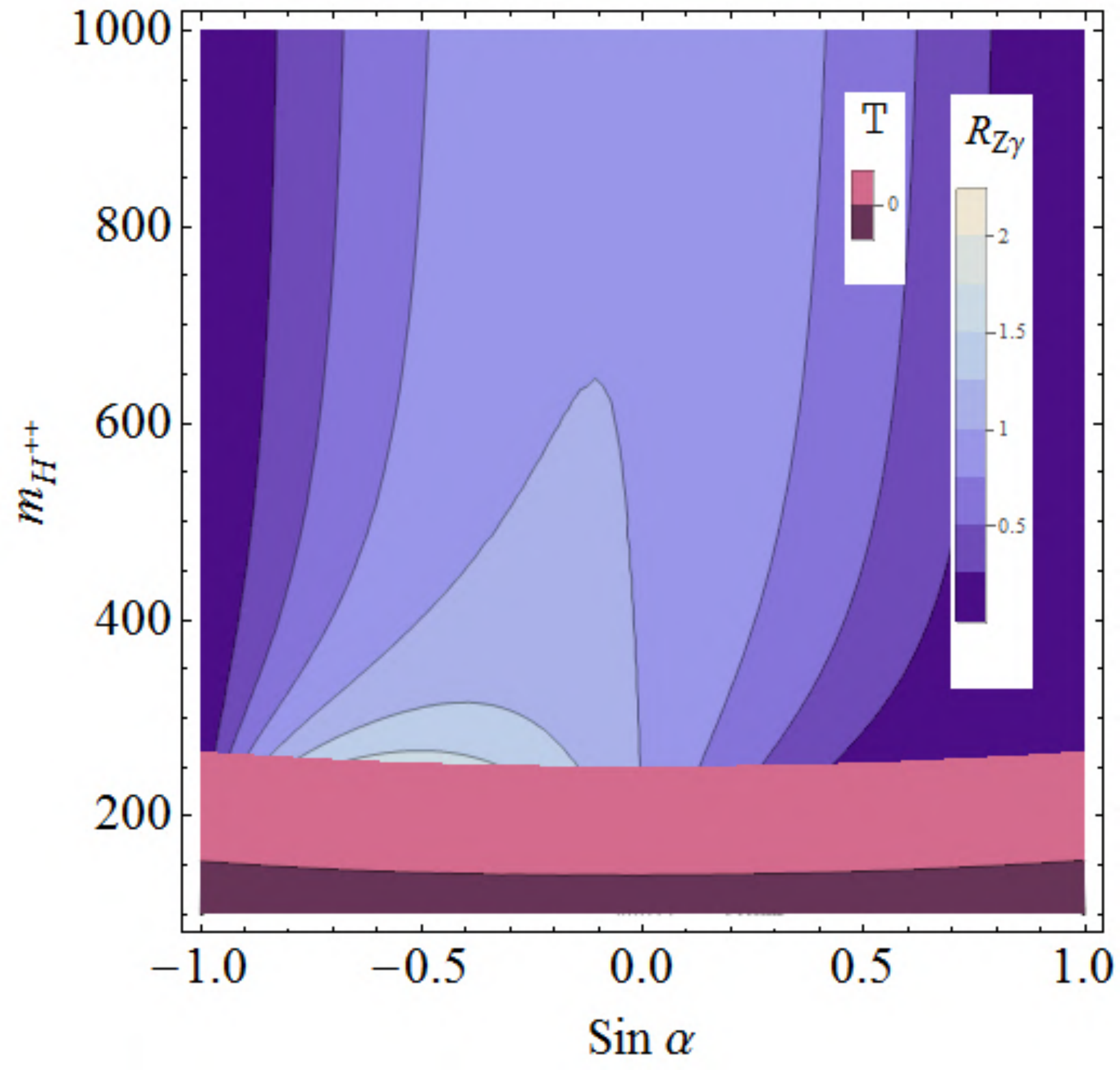}\\
        \end{array}$
\end{center}
\vskip -0.1in
     \caption{(color online). \sl\small Contour graphs for the relative strength of the decay of the Higgs boson into a photon an a Z-boson,  $R_{Z \gamma}$, including the restrictions of the $T$  parameter in the HTM with vector quarks, as a function of the doubly charged Higgs boson mass $m_{H^{\pm \pm}}$ and the mixing in the CP-even neutral Higgs sector, $\sin \alpha$. We show plots for scenario ${\cal D}_1$ (upper left panel), scenario ${\cal D}_X$ (upper middle panel),  scenario ${\cal D}_Y$ (upper right panel), scenario ${\cal T}_X$ (lower left panel), scenario ${\cal D}_2$ (middle lower panel) and scenario ${\cal U}_1$ (lower right panel).  Results for scenario ${\cal T}_Y$ (not shown) are summarized in the text. We choose the same parameters as in Fig. \ref{fig:Rgg}.}
\label{fig:RZg}
\end{figure}

The decay rates for $R_{Z \gamma}$ depend on $\sin \alpha$ and $m_{H^{\pm\pm}}$, though the variation is much milder than that for  $R_{\gamma \gamma}$. We investigate the dependence in Fig. \ref{fig:RZg} on the parameter space of $\sin \alpha$ - $m_{H^{\pm\pm}}$ through restrictions on the $T$ parameter. Contour plots for the $T$ parameter restrictions are shown in the (almost) horizontal bands, while values for $R_{Z \gamma}$ are shown in purple contours. Scales for both are included on the right panels. We have drawn plots for the same scenarios and the same order as for $R_{\gamma \gamma}$: ${\cal D}_1$, ${\cal D}_X$, ${\cal D}_Y$, ${\cal T}_X$, ${\cal D}_2$,  ${\cal U}_1$.   The features for $R_{Z\gamma}$ resemble those for $R_{\gamma \gamma}$. Distinguishing signs among  models come from restrictions on the values of the doubly charged Higgs boson mass. The relative branching ratios $R_{Z \gamma}$ are also sensitive to values of $\sin \alpha$.  For model ${\cal D}_1$,  in the regions allowed by $m_{H^{\pm \pm}}$ bands, the enhancement in the $Z\gamma$ decay can be at most  the SM value for  $\sin \alpha \in (-0.7, -0.02)$  and reaches $1.25$ when $\sin \alpha \in (-0.7, -0.02)$.  In model ${\cal D}_X$ also no enhancement is obtained for any values of the mixing angle and $R_{Z \gamma}=1$ for $\sin \alpha \in (-0.64, -0.02)$ region.   In the ${\cal D}_2$ model  enhancements of $R_{Z \gamma}$ of  $1.25-2$  are possible for both  negative  $\sin \alpha \in (-1, -0.12)$ values, and the same holds for the other models, ${\cal D}_Y$, ${\cal T}_X$,  ${\cal U}_1$ and ${\cal T}_Y$. Decays into $Z\gamma $ are correlated to those into $\gamma\gamma$ --that is, they are likely to be larger in the same regions of the parameter space, and for low doubly charged Higgs boson masses.


\clearpage

\section{Summary and Conclusion}
\label{sec:conclusion}

In this work we analyzed the effects of introducing vector-like states 
in the Higgs Triplet Model, allowed to be U-type or D-type  singlets (${\cal U}_1$, ${\cal D}_1$), SM-like or non SM, U-type or D-type doublets (${\cal D}_2$, ${\cal D}_X$, and ${\cal D}_Y$), and U-type or D-type triplets (${\cal T}_X$ and ${\cal T}_Y$). To conserve flavor, the only restriction we imposed was weak mixing with only the third family of ordinary quarks.

We posed the question: how does the introduction of these states affect the electroweak precision variables of the HTM. We were particularly interested in constraints on the mixing of the CP-even neutral Higgs bosons, the masses of the vector-like quarks and the mixing parameters with the ordinary quarks; and the mass of the doubly charged Higgs boson. We review here the constraints obtained in order.

First, the oblique parameters $S$, $T$ and $U$ were not all equally sensitive to mass parameters. We concentrated on the $T$ parameter, which showed significant variations with the doubly charged Higgs mass, in the absence of vector-like fermions. The doubly charged mass was restricted in a band around (100-280) GeV, varying very slightly with the triplet VEV, and about 10\% with the mixing angle in the neutral Higgs sector. And the contributions of the HTM model to the $T$ parameter ware found to be always positive.  
Addition of vector-like quarks also affects the $T$ parameter. While in models ${\cal U}_1$, ${\cal D}_2$, ${\cal D}_Y$, ${\cal T}_X$ and  ${\cal T}_Y$ their contribution is always positive, in models ${\cal D}_1$, ${\cal D}_X$ , there is a region of parameter space where the contribution is negative, thus subtracting from the contribution from the doubly charged Higgs bosons and raising the bound on their masses. We have investigated this in detail for models ${\cal D}_1$ and ${\cal D}_X$, as the negative contribution to the $T$ parameter occurs for a larger range of the mixing parameter $x_b$ and $x_t$.  The ${\cal D}_1$ and ${\cal D}_X$ models are then  distinguishable in this framework, as they require doubly charged Higgs boson masses in the $\sim 300-400$ GeV region to satisfy electroweak constraints, while the other models require significantly lighter doubly charged Higgs bosons in the $\sim 100 - 290$ GeV region.  Electroweak precision data also restricts the mixing parameters $x_b$ in the $(117-400)$ GeV range for relatively light vector-like quark masses ($M \sim 300$ GeV), while for heavier masses, $M \sim 1000$ GeV, the mixing parameters range is increased to $x_b \in (117-550)$ GeV. The lower limit comes from  $Zb {\bar b}$ constraints. A different restriction occurs for $x_t$. First, the $Wtb$ vertex does not impose a lower limit, and second, the range of this parameter decreases when the mass of the vector-like quark mass increases, so that  $x_t  \in (0, 725)$ GeV for $M \sim 300$ GeV, and $x_t  \in (0, 550)$ GeV for $M \sim 1000$ GeV.

The effects of vector-like parameters on limits on the doubly charged Higgs boson masses are as follows. In the models  ${\cal U}_1$, ${\cal D}_2$, ${\cal D}_Y$, ${\cal T}_X$,  increasing $M$ and decreasing $x_{t(b)}$ yields a slightly higher upper band for  doubly charged Higgs bosons mass. While in the ${\cal T}_Y$ model, where very light $x_t$ values are required, decreasing these mixing parameters increases the doubly charged mass,  and in ${\cal  D}_1$ and ${\cal D}_X$ models,  higher upper limits for $m_{H{\pm \pm}}$ are obtained for lighter vector-like quark masses $M$.

While the production and decay mechanisms of the vector-like quarks are not modified by the particles in the HTM (as the only new particles, the triplet Higgs bosons, do not couple to quarks), loop-induced decays of the neutral Higgs bosons are affected. Interestingly, while the masses and mixing parameters of the vector-like quarks have little effect on the $H \to \gamma \gamma$ and $H \to Z \gamma$ decays, as in the SM, the effects of vector-like quarks come from combining these with constraints from electroweak precision observables. 

These observables restrict the doubly charged Higgs boson mass to be in the (about) 300-400 GeV for models ${\cal D}_1$ and ${\cal D}_X$, and (about) 100-290 GeV for the rest of the models. Enhancement of the rates  $R_{\gamma \gamma}$ and $R_{Z \gamma}$  are more likely to occur at negative values of $\sin \alpha$, the mixing angle in the neutral Higgs sector, and in particular, for model ${\cal D}_X$, only for negative values.  Scenarios ${\cal D}_1$ and  ${\cal D}_X$ scenario predict no enhancement of $R_{Z \gamma}$ in the allowed parameter region. Thus in the HTM, scenarios ${\cal D}_1$ and ${\cal D}_X$ stand out as distinguishable from the rest (from doubly charged Higgs boson mass restrictions) and from each other (from regions and strength of possible enhancements in loop dominated Higgs decays). 

To summarize, introducing vector-like quarks in the Higgs Triplet Model alters the electroweak constraints on the parameters of the model and yields tighter predictions for the enhancement of loop-dominated Higgs decays, expected to be measured even more precisely at the LHC operating at 13 TeV.


\section{Appendix}
\label{sec:appendix}
\appendix

We list below the $W$ and $Z$ couplings in quark-like models used to restrict masses and mixings in the ${\cal D}_1$ and ${\cal D}_X$ models.
\begin{table}[htbp]
\caption{\label{tab:llW} Couplings to the $W$ and $Z$ bosons}
  \begin{center}
 \small
 \begin{tabular*}{0.99\textwidth}{@{\extracolsep{\fill}} c ccccc}
 \hline\hline
 {\sl \small Light-light couplings to the W boson}\\
 \hline\hline
	{\sl\small model / matrix element} &$V_{tb}^L$ &&$V_{tb}^R$ \\
  \hline\hline
  ${\cal D}_1$ &$c_L^{d}$ &&$0$ \\
  ${\cal D}_X$ &$c_L^{u}$& &$0$ \\
    \hline
    \hline
    {\sl\small Heavy-heavy couplings to the W boson}\\
 \hline\hline
	{\sl\small model / matrix element} &$V_{XT}^L$ &&$V_{XT}^R$ 
	\\
  \hline\hline
  ${\cal D}_X$ &$c_L^{u}$ &&$c_R^{u}$ 
    \\
    \hline
    \hline
     {\sl\small Heavy-light couplings to the W boson}\\
 \hline\hline
	{\sl\small model / matrix element} &$V_{Xt}^L$ &$V_{Xt}^R$ &$V_{Tb}^L$ &$V_{Tb}^R$  \\
  \hline\hline
  ${\cal D}_X$ &$-s_L^{u}$ &$-s_R^{u}$ &$s_L^{u}$ &$0$ \\
     \hline
    \hline
     {\sl\small Light-heavy couplings to the W boson}\\
 \hline\hline
	{\sl\small model / matrix element} &$V_{tB}^L$ &&$V_{tB}^R$   
	\\
  \hline\hline
	${\cal D}_1$ &$s_L^{d}$ &&$0$   \\
    \hline
    \hline
     {\sl\small Light-heavy couplings to the Z boson}\\
 \hline\hline
	{\sl\small model / matrix element} &$U_{tT}^L$ &$U_{tT}^R$ &$D_{bB}^L$ &$D_{bB}^R$\\
  \hline\hline
    ${\cal D}_1$ &$ $ &$ $ &$s_L^{d}c_L^{d}$ &$0$ \\
  ${\cal D}_X$ &$2s_L^{u}c_L^{u}$ &$s_R^{u}c_R^{u}$ &$ $ &$ $ \\
    \hline
    \hline
   \end{tabular*}
\end{center}
 \end{table}
\clearpage
We list below the Higgs boson couplings in quark-like models.
\begin{table}[htbp]
\caption{\label{tab:llh} Couplings to the Higgs bosons}
  \begin{center}
 \small
 \begin{tabular*}{0.99\textwidth}{@{\extracolsep{\fill}} c ccccc}
 \hline\hline
 {\sl \small Light-light couplings to the Higgs boson}\\
 \hline\hline
	{\sl\small model / matrix element} &$Y_{tt}$ &&$Y_{bb}$ \\
  \hline\hline
  ${\cal U}_1$ &${c_L^{u}}^2$ &&$1$ \\
  ${\cal D}_1$ &$1$ &&${c_L^{d}}^2$ \\
  ${\cal D}_X$ &${c_R^{u}}^2$& &$1$ \\
	${\cal D}_2$ &${c_R^{u}}^2$& &${c_R^{d}}^2$ \\
	${\cal D}_Y$ &$1$& &${c_R^{d}}^2$ \\
	${\cal T}_X$ &${c_L^{u}}^2$& &${c_L^{d}}^2$ \\
	${\cal T}_Y$ &${c_L^{u}}^2$& &${c_L^{d}}^2$ \\
    \hline
    \hline
    {\sl\small Heavy-heavy couplings to the Higgs boson}\\
 \hline\hline
	{\sl\small model / matrix element} &$Y_{TT}$ &&$Y_{BB}$ 
	\\
  \hline\hline
  ${\cal U}_1$ &${s_L^{u}}^2$ &&$ $ \\
  ${\cal D}_1$ &$ $ &&${s_L^{d}}^2$ \\
  ${\cal D}_X$ &${s_R^{u}}^2$& &$ $ \\
	${\cal D}_2$ &${s_R^{u}}^2$& &${s_R^{d}}^2$ \\
	${\cal D}_Y$ &$ $& &${s_R^{d}}^2$ \\
	${\cal T}_X$ &${s_L^{u}}^2$& &${s_L^{d}}^2$ \\
	${\cal T}_Y$ &${s_L^{u}}^2$& &${s_L^{d}}^2$ \\
  \\
    \hline
    \hline
		\end{tabular*}
\end{center}
 \end{table}

\clearpage

\acknowledgments
The work of S. B. and M.F.  is supported in part by NSERC under grant number SAP105354.


\end{document}